\documentclass[twocolumn,preprintnumbers,amsmath,amssymb,aps,pra,showpacs,floatfix,superscriptaddress]{revtex4-1}
\usepackage{natbib}
\usepackage{graphicx}
\usepackage{epsfig}
\usepackage{dcolumn}
\usepackage{bm}
\usepackage{soul}
\usepackage{sidecap}
\usepackage[usenames,dvipsnames]{color}
\usepackage{setspace}
\usepackage[colorlinks=true,linkcolor=blue,urlcolor=blue,citecolor=blue]{hyperref}
\usepackage[usenames,dvipsnames,svgnames,table]{xcolor}
\usepackage{wasysym}
\usepackage{multirow}

\begin{document}
\title{Heavy non-degenerate electrons in doped strontium titanate}

\author{Cl\'ement Collignon}
\altaffiliation[Present address: ]{Department of Physics, Massachusetts Institute of Technology, Cambridge, Massachusetts
02139, USA}
\affiliation{JEIP, USR 3573 CNRS, Coll\`ege de France, PSL University, 11, place Marcelin Berthelot, 75231 Paris Cedex 05, France }
\affiliation{Laboratoire Physique et Etude de Mat\'{e}riaux (CNRS-Sorbonne Universit\'e), ESPCI, PSL Research University, 75005, Paris, France}

\author{Phillipe Bourges}
\affiliation{Laboratoire L\'eon Brillouin, CEA-CNRS, Universit\'e Paris-Saclay, CEA Saclay, 91191 Gif-sur-Yvette, France}

\author{Beno\^{\i}t Fauqu\'e}
\email{benoit.fauque@espci.fr}
\affiliation{JEIP, USR 3573 CNRS, Coll\`ege de France, PSL University, 11, place Marcelin Berthelot, 75231 Paris Cedex 05, France }

\author{Kamran Behnia$^{2}$} 

\date{\today}

\begin{abstract}
Room-temperature metallicity of lightly doped SrTiO$_3$ is puzzling, because the combination of mobility and the effective mass would imply a mean-free-path (mfp) below the Mott Ioffe Regel (MIR) limit and a scattering time shorter than the Planckian time ($\tau_P=\hbar/k_BT$). We present a study of electric resistivity, Seebeck coefficient and inelastic neutron scattering extended to very high temperatures, which deepens the puzzle. Metallic resistivity persists up to 900 K and is accompanied by a large Seebeck coefficient whose magnitude (as well as its temperature and doping dependence) indicates that carriers are becoming heavier with rising temperature.  Combining this with neutron scattering data, we find that between 500 K and 900 K, the Bohr radius and the electron wave-length become comparable to each other and twice the lattice parameter. According to our results, between 100 K and  500 K, metallicity is partially driven by temperature-induced amplification of the carrier mass. We contrast this mass amplification of non-degenerate electrons with the better-known case of heavy degenerate electrons. Above 500 K,   the mean-free-path continues to shrink with warming in spite of becoming shorter than both the interatomic distance and the thermal wavelength of the electrons. The latter saturates to twice the lattice parameter. Available theories of polaronic quasi-particles do not provide satisfactory explanation for our observations.
\end{abstract}
\maketitle

\section{Introduction}

Decades ago~\cite{Mott1956}, Mott argued that the threshold of metallicity in a doped semiconductor depends on the effective range of the Coulomb interaction exerted by an extrinsic atom, the Bohr radius. According to what is now known as the Mott criterion for metal-insulator transition~\cite{Edwards1978,Mott1990}, a semiconductor becomes metallic when the density of its carriers, $n$ exceeds a threshold set by its effective Bohr radius, $a_B^{\star}$ : 
\begin{equation}
n^{1/3}a_B^{\star} >0.25
\label{Mott}
\end{equation}
Since $a_B^{\star} = 4 \pi \epsilon \hbar / m^{\star} e^2$ (where $m^{\star}$ is the mass carrier and $\epsilon$ the dielectric constant), a small $m^{\star}$ or a large $\epsilon$ would favor precocious metallicity when a semiconductor is doped. 

Strontium titanate is a wide-gap semiconductor whose electric permittivity becomes as large as 20$\,$000 times the vacuum permittivity in liquid helium temperature~\cite{Muller1979,Hemberger1995,Sirenko2000,Rossle2013}, which implies a Bohr radius approaching a micron~\cite{behnia2015mobility}. As a consequence, one can easily turn it to a metal~\cite{Spinelli2010} with one carrier per 10$^5$ unit cells. This dilute metal has attracted renewed attention in recent years for multiple reasons~\cite{collignon2019metallicity}. First of all, it becomes a superconductor~\cite{Gastiasoro2020}, implying that Cooper pairs can be formed even when the Fermi energy is an order of magnitude lower than the Debye energy~\cite{Lin2013,Edge2015,Ruhman2016,Gorkov2016,wolfle2018,vandermarel2019,Klimin2019}. The dilute superconductor~\cite{Lin2014_multiple,Lin2015} appears to be intimately linked to aborted ferroelectricity~\cite{Rowley2014,Stucky2016,Rischau2017,Tomioka2019,Ahadi2019,Herrera2019}. Second, the high mobility of carriers at low temperature allows the observation of quantum oscillations~\cite{Lin2013,Allen2013,Lin2014,Bhattacharya2016} in relatively low magnetic fields. The evolution of the Fermi surface with doping can be explored and compared with what is expected by theory~\cite{vdMarel2011}, and test the limits of rigid-band approximation. Third, the low-temperature resistivity displays a quadratic temperature dependence~\cite{Tokura1993,vdMarel2011,Lin2015sc,mikheev2015,mikheev2016,Wang2019} with a prefactor which smoothly increases with decreasing carrier concentration~\cite{Lin2015sc}. Such a $T$-square resistivity is expected in a Fermi liquid with dominant e$^-$- e$^-$ scattering. However, here, the behavior persists even in the extreme dilute limit in absence of Umklapp scattering~\cite{Lin2015sc}. Finally, the high-temperature metallicity is 'beyond quasi-particles'~\cite{Lin2017,mishchenko2019,Zhou2019}. The combination of room-temperature resistivity and low-temperature effective mass implies a mean-free-path that falls below all known length scales of the solid (the electron wavelength and the lattice parameter)~\cite{Lin2017}. This has been observed in strange metals with strong correlation among electrons~\cite{Gunnarsson2003}, in organic semiconductors \cite{Fratini2016}, but not in inorganic doped band insulators.

In this paper, we will address this last issue by measuring the resistivity and the Seebeck coefficient of SrTi$_{1-x}$Nb$_x$O$_3$ up to temperatures as high as 900 K, in which, in contrast to oxygen-reduced strontium titanate, exposition to high temperatures does not modify the number of dopants (see methods). We find that even at 900 K, resistivity continues to increase and the Seebeck data implies that electron mass evolves as a function of temperature. We also present a study of inelastic neutron scattering, which documents the evolution of the soft zone center transverse optical (TO) phonon mode above room temperature. After presenting our resistivity data, we will demonstrate the features shared by metallic strontium titanate with other systems close to a ferroelectric instability in contrast to ordinary metallic (i.e. 'heavily doped' \cite{Shklovskii1984}) semiconductors. We will see that the magnitude of mobility combined with low-temperature effective mass would imply a scattering time shorter than the Planckian time, a feature which would distinguish this metal from other 'strange' metals, which respect this limit~\cite{bruin2013}. Then we will present our Seebeck data and argue that the expression for the thermoelectric response of non-degenerate electrons~\cite{johnson1953} implies a temperature-induced amplification of the carrier mass. The extracted heavy masses, shift upward the temperature window where the scattering time falls below the Planckian time. The amplified mass combined with the dielectric permittivity extracted from the neutron data allows us to conclude that for $T>500$ K, the Bohr radius and the thermal wave-length both shrink to twice the lattice parameter and much shorter than the interelectron distance. The persistence of metallicity in this context remains beyond any available quasiparticle-based picture and a new challenge to theory, which has started to tackle charge transport in dilute metallic strontium titanate~\cite{mishchenko2019,Zhou2019}.

One conclusion is that in this metal, the temperature dependence of resistivity is partially set by the evolution of the effective mass. This idea was previously put forward by Eagles, who invoked 'mixed polarons'~\cite{Eagles1996}. Nevertheless, we will argue that the underlying microscopic interaction is yet to be identified.

\section{Results}
\subsection{Resistivity of doped strontium titanate from 2 K to 900 K}

\begin{figure}[h!]
\centering
\includegraphics[width=\linewidth]{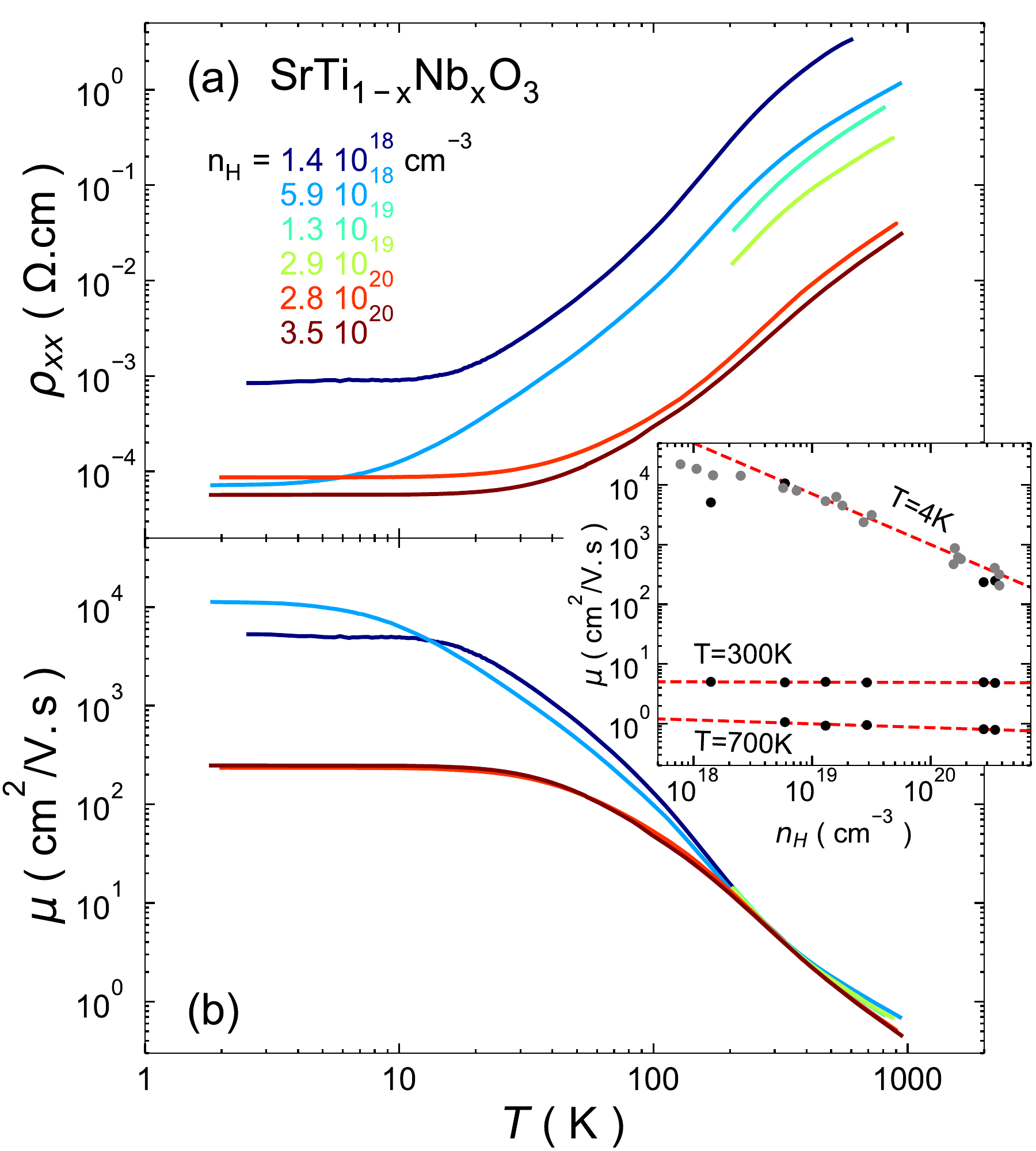}
\caption{\textbf{a)}: Temperature dependency of the resistivity extended to 900 K for several Nb doped SrTiO$_3$ crystals. The metallic behavior continues up to 900 K.
\textbf{b)}: Temperature dependency of the Hall mobility $\mu = 1 /(\rho \, n_H \, e)$ from 2 to 900 K of the same samples. The inset shows the dependence of mobility on carrier concentration at different temperatures. At room temperature and above the mobility is low and show little variation with carrier concentration. At low temperatures, the mobility is large and strongly depends on carrier concentration.}
\label{Resistivity}
\end{figure}

Fig. \ref{Resistivity}(a) shows the temperature dependence of resistivity for six niobium doped samples with carrier concentrations ranging from 1.4 $\times 10^{18}$ to 3.5 $\times 10^{20}$ cm$^{-3}$. Our data below 300 K is in agreement with previous old~\cite{frederikse1964,frederikse1967,Tufte1967} and recent~\cite{ Spinelli2010,cain2013,Lin2015sc,mikheev2015,Lin2017} studies of charge transport in this system (see ref. \cite{collignon2019metallicity} for a review). Let us note that reproducible measurements at high temperatures are challenging, because of the possible variation in the number of oxygen vacancies with increasing temperature. Our measurements were performed in presence of adequate air pressure. Frederikse and Hosler~\cite{frederikse1967} checked that the Hall number, $n_H$ of Nb-doped strontium titanate remains temperature-independent up to 1000 K.

As one can see in the figure, the metallic resistivity, problematic even at room temperature~\cite{Lin2017}, persists up to 900 K. Fig. \ref{Resistivity}(b) displays the temperature dependence of mobility. Upon warming, it changes by four orders of magnitude. The room-temperature mobility is as low as $5$ cm$^2$/V.s. As we will see below, this is uncommonly small among metallic semiconductors. Upon warming to 900 K, it falls below $1$ cm$^2$/V.s. As shown in the inset of Fig. \ref{Resistivity}, at 900 K, with a hundred fold increase of carrier concentration, mobility slightly decreases from 0.7 to 0.5 cm$^{2}$/V.s. At room temperature, it does not display any detectable dependence on carrier concentration. In contrast, at low temperature, when impurity scattering dominates, it becomes orders of magnitude larger and shows a strong dependence on carrier density. As we will see below, the presence or absence of these features distintiguish two groups of dilute metals.


\subsection{The mean-free-path and the scattering time}
\begin{figure}[h!]
\centering
\includegraphics[width=\linewidth]{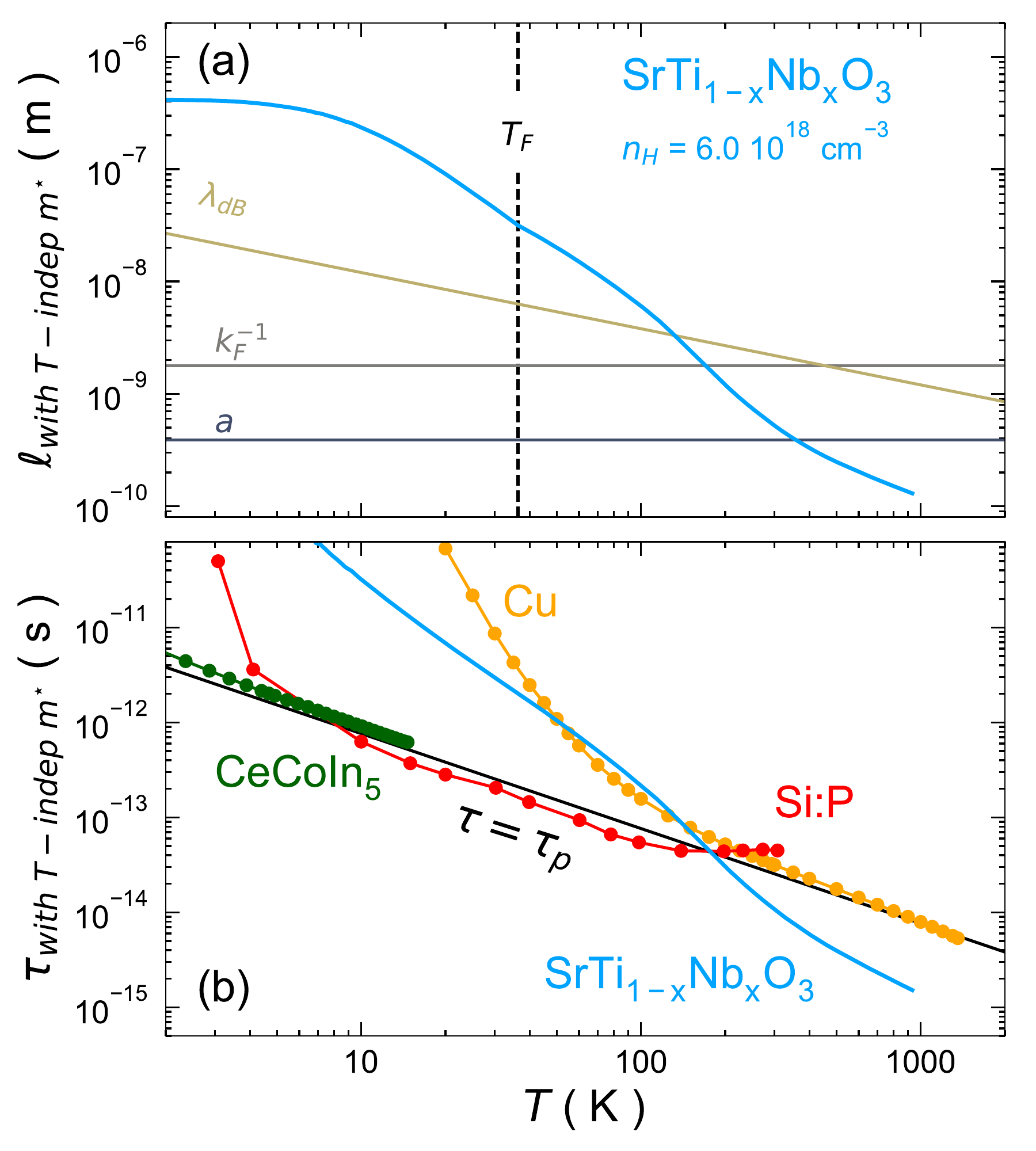}
\caption{
\textbf{a)}: Temperature dependency of the mean free path, $\ell$, of Nb doped SrTiO$_3$ for $n_H = 5.9$ 10$^{18}$ cm$^{-3}$, using the low-temperature effective mass.
The lattice parameter, $a$, the inverse of Fermi wave-vector, $k_F^{-1}$, and the de Broglie thermal wavelength $\lambda_{dB}$ are also plotted. $\ell$ is computed using Eq. \ref{mfp_D} when $T < T_F$ and using Eq. \ref{mfp_ND} when $T>T_F$. The mass used to compute $\ell$ and $\lambda_{dB}$ is the one measured at low temperature through quantum oscillations \cite{Lin2014}.
\textbf{b)}: Temperature dependency of the inelastic scattering time, $\tau=m^{\star}/ne^2(\rho-\rho_0)$, of the same sample. Also shown are the inelastic scattering time of P-doped Si (a doped semiconductor), copper (a good metal) and CeCoIn$_5$ (a bad metal). The Planckian time $\hbar / k_B T$ is also plotted in black.
Except for Nb doped SrTiO$_3$, the scattering time in all cases remain longer than the Planckian time. Note that electrons in Cu and in CeCoIn$_5$ remain degenerate, but in P- doped Si and Nb doped SrTiO$_3$, they become non- degenerate above 60 K and 40 K respectively. The resistivity and the effective mass data for Si, Cu and CeCoIn$_5$ are taken from ref. \cite{Yamanouchi1967,barber1967,Koch1964,Matula1979,Settai2001,McCollam2005,Tanatar2007}.
}
\label{MIR_Pt}
\end{figure}

As previously noticed~\cite{Lin2017}, the measured mobility of $\mu = 4.9 \pm 0.5$ cm$^{2}$/V.s at $T=300$ K implies a mean-free-path falling below any relevant length scale of the system. Since the mobility becomes almost an order of magnitude smaller at 900 K, the breakdown of the quasi-particle picture is becoming more drastic. This can be seen in Fig. \ref{MIR_Pt}(a) which shows the temperature dependence of the mean free path of the sample with $n_H =5.9 \times$ 10$^{18}$ cm$^{-3}$. Also plotted are the inverse Fermi wave vector $k_F^{-1}$, the lattice parameter $a$ and the de Broglie wavelentgh $\lambda_{dB}=\sqrt{2\pi \hbar^2/m^{\star}k_B T}$.

Below the degeneracy temperature, $T_F = 39$ K the velocity of electrons is $v_F=\hbar k_F / m^{\star}$. Therefore, assuming a simple Drude law, $\sigma=ne^2\tau / m^{\star}$, and by noting that $\ell=v_F \, \tau$ and $k_F^3=3\pi^2n$ we get a mass independent mean free path:
\begin{equation}
\ell_{T<T_F} =\frac{\mu}{e} \, \hbar k_F=\frac{\mu}{e} \hbar (3 \pi^2 n)^{1/3}
\label{mfp_D}
\end{equation}
where $\rho$ is the resistivity and $n$ the carrier concentration.

Above $T_F$, electrons become non-degenerate and the velocity of carriers is not the Fermi velocity but the thermal velocity $v_{th} = \sqrt{2k_BT/m^{\star}}$. The scattering time can be extracted from resistivity, assuming once again a simple Drude law and $\ell=v_{th} \, \tau$:
\begin{equation}
 \ell_{T>T_F} = (2\sqrt{\pi}) \, \, \frac{\mu}{e} \, \hbar\lambda_{dB}^{-1}=\frac{\mu}{e}\sqrt{2m^{\star}k_{B}T}
\label{mfp_ND}
\end{equation}

Note that in both cases mobility is simply the ratio of the mean-free-path to momentum. The only difference is that this momentum is the Fermi momentum below $T_F$ and the thermal momentum above. According to Eq. \ref{mfp_D}, one needs the effective mass to quantify $\lambda_{dB}$ and $\ell$ in the non-degenerate regime. In Fig. \ref{MIR_Pt}, it has been assumed that $m^{\star} \simeq 3.8 ~m_e$, which is the heaviest cyclotron mass detected by quantum oscillations at low temperatures \cite{Lin2014}. Taking a lighter mass would shorten further the mean free path. The figure shows that at 900 K, the mean free path becomes four times shorter than the inter-lattice spacing and one order of magnitude below $\lambda_{dB}$.

The so-called Mott-Ioffe-Regel (MIR) limit~\cite{Ioffe1960, Mott1972} is a lower boundary to the mean free path of carriers of a metal. This minimum length is either the lattice parameter or the quasi-particle wavelength~\cite{Emery1995}. Most metals respect this limit and their resistivity saturates when the mean-free-path becomes too short~\cite{mooij1973,Fisk1976}. Metals, which do not respect this limit~\cite{Gunnarson2003,Hussey2004} were dubbed bad metals~\cite{Emery1995}. Bruin and co-workers~\cite{bruin2013} noticed that good and bad metals both show a scattering time which does not fall below the Planckian time ($\tau_p=\hbar / k_B T$). This led to a theoretical proposal for the existence of such a bound to diffusive transport based on Heisenberg uncertainty~\cite{Hartnoll2015}.
 
The scattering time, $\tau$, of our system can be easily quantified. It is the velocity (thermal above $T_F$ and Fermi below) divided by the mean-free-path. As seen in the lower panel of Fig. \ref{MIR_Pt}, at 900 K, $\tau$ falls one order of magnitude below $\tau_p$. For comparison, we computed the inelastic scattering time of a good metal (pure Cu), a bad metal (CeCoIn$_5$) and a doped semiconductor (P doped Si) using their resistivity and effective mass, $\tau=m^{\star}/ne^2(\rho-\rho_0)$, extracted from published data~\cite{Yamanouchi1967,barber1967,Koch1964,Matula1979,Settai2001,McCollam2005,Tanatar2007}. As seen in the figure, their scattering time does not become shorter than the Planckian time as noticed by Bruin and co-workers~\cite{bruin2013}.

At this stage, doped strontium titanate appears to behave remarkably bad, worse than cuprates, the most notorious of bad metals, which have been shown to obey the Planckian bound~\cite{Legros2018}, even though they do not respect the MIR limit~\cite{Gunnarson2003,Hussey2004}.

\subsection{Two distinct types of dilute metals}
\begin{figure}[h!]
\centering
\includegraphics[width=\linewidth]{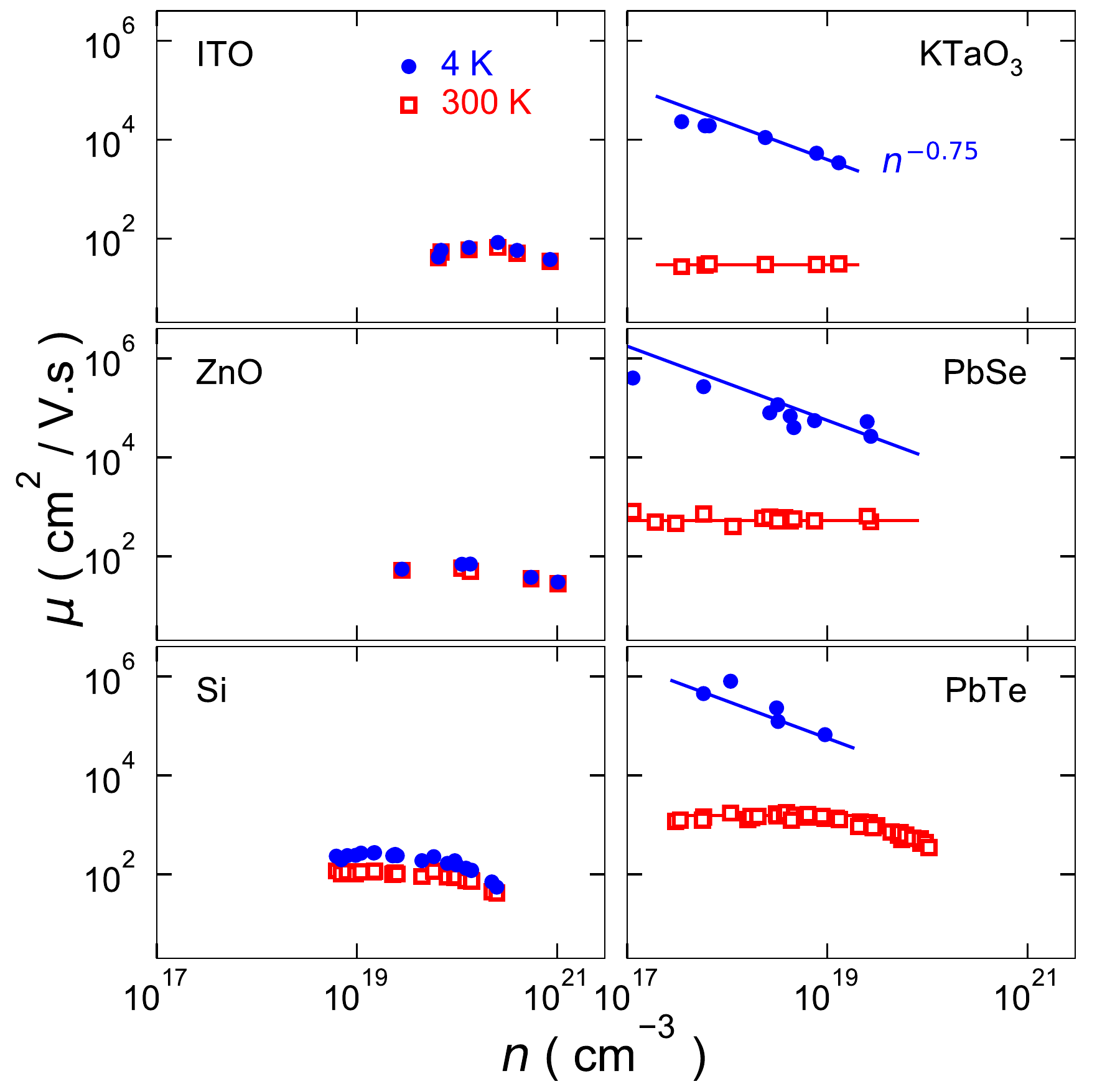}
\caption{Mobility as a function of doping at $T = 4 $ K (full circles) and $T = 300 $ K (open squares) in different dilute metals. Right panels show the data for tin-doped In$_2$O$_3$ (ITO), Al and Ga doped ZnO and phosphorus-doped silicon~\cite{huang1987,das2012,ahn2007,naidu2012,Yamanouchi1967}. Left panels show the data for KTaO$_3$, PbTe, and PbSe~\cite{engelmayer2019,wemple1965,Allgaier1958}, which are all polar semiconductors close to a ferroelectric instability. In the first case, mobility does not change much with cooling or variation of carrier density. In the second case, mobility at cryogenic temperatures is much larger than at room temperature and displays a distinct power low dependence on mobility $\mu(4$ K$)\propto n^{-\alpha}$, with $\alpha \simeq0.75 $}
\label{Mobility_300&4K}
\end{figure}

It is surprising to find such a strangeness in what is, after all, merely a doped band insulator. Let us compare our system to other doped semiconductors. The origin of the absurdly short scattering time of strontium titanate is the large temperature dependence of mobility. It changes by four orders of magnitude between 900 K and 2 K [see Fig. \ref{Resistivity}(b)]. Moreover, the 2 K mobility varies by a factor 200 between $n_H = 6\times 10^{18}$ and $3 \times 10^{20}$ cm$^{-3}$ (the inset of the same figure). 

Neither of these two features can be seen in ordinary semiconductors. In metallic phosphorous-doped silicon \cite{Yamanouchi1967}, in Sn doped indium-oxide (ITO) \cite{huang1987} and in doped ZnO~\cite{ahn2007,das2012,naidu2012}, mobility displays little change with temperature and a modest decrease with the increase in carrier concentration (see Fig. \ref{Mobility_300&4K}). On the other hand, other polar semiconductors, which are close to a ferroelectric instability, such as IV-VI semiconductors (PbTe and PbSe) and ABO$_3$ perovskytes (like KTaO$_3$ and EuTiO$_3$) display the two features seen in doped strontium titanate. Specifically, in these systems, mobility enhances by orders of magnitude upon cooling, and in cryogenic temperatures, displays a power-law dependence on carrier concentration (see Fig. \ref{Mobility_300&4K}). It has been argued~\cite{behnia2015mobility}, that the latter feature is an expected consequence of a Bohr radius exceeding by far the interatomic distance.

Like strontium titanate, these solids are quantum paraelectrics with soft phonons. However, since their room-temperature mobility is larger than strontium titanate, their mean-free-path and their scattering time remain reasonably long. Nevertheless, the qualitative similarity seen among these dilute metals and the contrast with ordinary doped semiconductors suggest that the presence of a soft ferroelectric mode plays a role in the peculiar metallicity of doped strontium titanate.

\subsection{High-temperature Seebeck coefficient}


\begin{figure*}
\centering
\includegraphics[width=18cm]{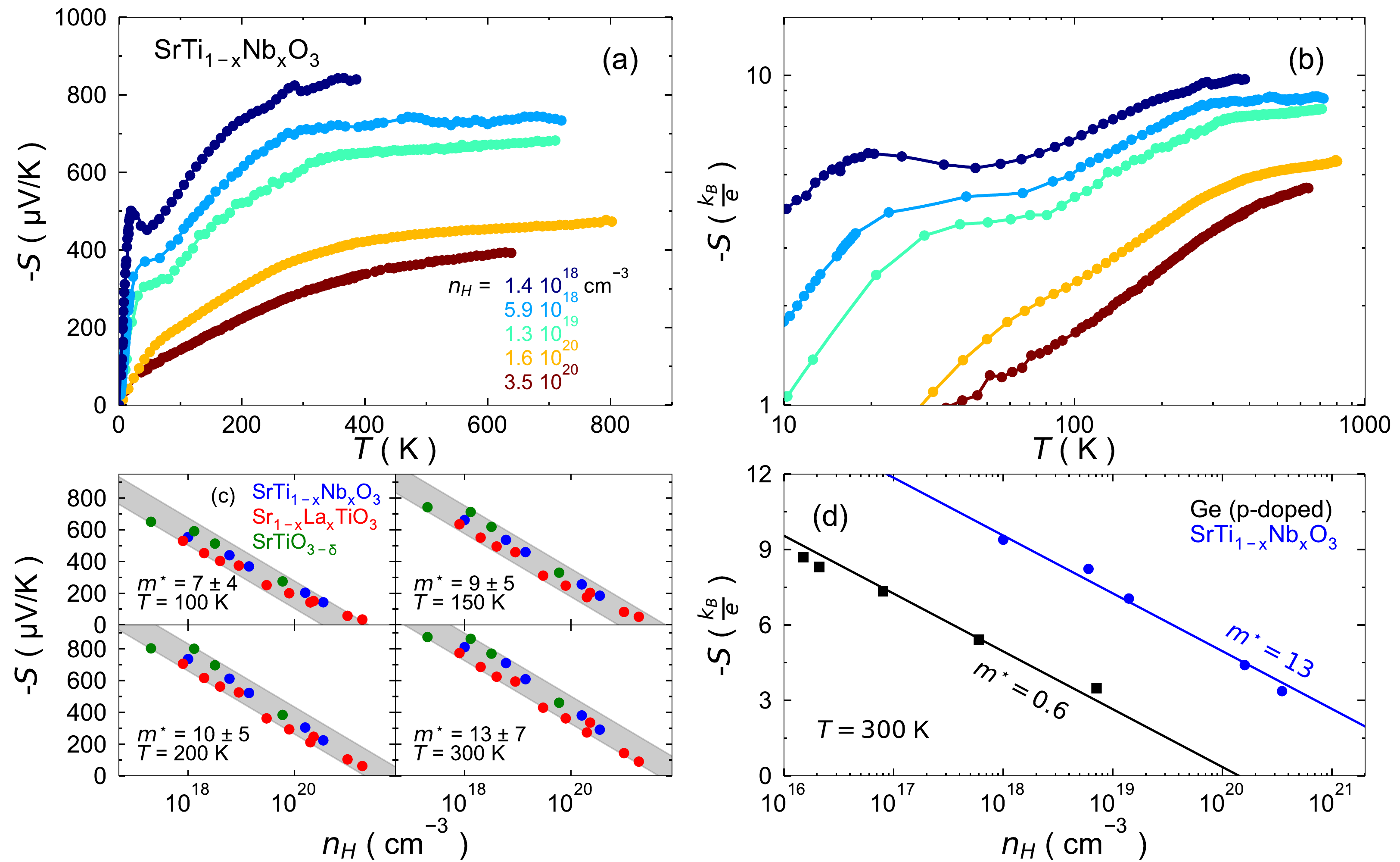}
\caption{\textbf{a)}: The Seebeck coefficient as a function of temperature for five Nb doped SrTiO$_3$ samples from 10 K to 800 K. \textbf{b)}: The same data in a log-log plot with $S$ expressed in units of $k_B/e$. Note the smooth evolution with doping, the presence of a low-temperature phonon drag peak. In the most dilute sample, the thermoelectric power becomes as large as 10 $k_B/e$, which implies a very large entropy per charge carrier. \textbf{c)}: Variation of the Seebeck coefficient with carrier concentration at different temperatures. Blue circles represent our Nb-doped samples, while green circles represent oxygen-reduced samples~\cite{frederikse1964} and red circles are La-doped samples~\cite{cain2013,Okuda2001}. Thick gray lines represents the behavior expected by Eq. \ref{Pisarenko}. The corresponding effective mass is indicated with an uncertainty set by the width of the line. \textbf{d):} A comparison of the room-temperature Seebeck coefficient, $S$ of p-type Ge \cite{Geballe1954} and Nb doped strontium titanate. In both systems $S$ follows a $-\log(n)$ dependence, as expected by Eq. \ref{Pisarenko}. However, at the same carrier concentration, non-degenerate carriers have more entropy in strontium titanate, implying that they are heavier.}
\label{Seebeck}
\end{figure*}

The temperature dependence of the Seebeck coefficient for five Nb doped SrTiO$_3$ samples between 10 K to 800 K is shown in figure \ref{Seebeck}(a). Our data is in reasonable agreement with previous reports below room temperature \cite{Okuda2001,Lin2013,cain2013}. Lin \textit{et al.}~\cite{Lin2013} reported that the Seebeck coefficient is $T$-linear at low temperature as expected for diffusive thermoelectric response of degenerate electrons ($S = \frac{k_B}{e} \frac{\pi^2}{3}\frac{T}{T_F}$ \cite{Behnia2015b}). Moreover, the low-temperature slope, $S/T$, was found to be in excellent agreement with the magnitude of the Fermi temperature extracted from quantum oscillations. Cain \textit{et al.} documented the evolution of the Seebeck coefficient of Sr$_{1-x}$La$_x$TiO$_3$ and found a phonon-drag peak around 25 K~\cite{cain2013}, which is also present in our data [Fig. \ref{Seebeck}(b)]. The fact that this peak occurs near the peak temperature of the lattice thermal conductivity~\cite{Martelli2018} supports the interpretation that the peak is caused by phonon drag as originally suggested~\cite{cain2013}. 

Our focus here is the high-temperature regime, well above both the degeneracy temperature and the phonon drag regime. In this temperature range, the magnitude of Seebeck coefficient is given by Eq. \ref{Pisarenko}, dubbed 'the Pisarenko formula' by Ioffe~\cite{Ioffe1957} (and many subsequent authors): 

\begin{equation}
 |S| = \frac{k_B}{e} \left[ 2+r + \ln \left( \frac{2}{n\Lambda^3} \right) \right]
\label{Pisarenko}
\end{equation}

We note that an equation identical to this was already derived by Johnson and Lark-Horovitz as the expression for the Seebeck coefficient of non-degenerate electrons in germanium crystals~\cite{johnson1953} when the carrier density is set by extrinsic dopants (and not by thermal excitation of carriers across the band gap). Here, $r$ (a.k.a the scattering parameter), represents the energy dependence of the scattering time $\tau \propto E^{r-1/2} $, and $\Lambda$ should be the de Broglie thermal wavelength $\lambda_{dB} = \sqrt{\frac{2\pi \hbar^2}{m^{\star}k_B T}}$. When the mean-free-path is independent of energy, then $r=0$, which is what Johnson and Lark-Horovitz~\cite{johnson1953} assumed in the case of germanium ~\cite{johnson1953}. A simple derivation of Eq. \ref{Pisarenko} is given in section III of the supplementary~\cite{SM}.

Let us note that there is a shortcut route towards Eq. \ref{Pisarenko} thanks to thermodynamics. There is indeed a fundamental link between Eq. \ref{Pisarenko} and the Sackur-Tetrode~\cite{Sackur1911,Sackur1912,tetrode1912} entropy, $\mathcal{S}_{ST}$, of a mono-atomic ideal gas of $N$ atoms~\cite{Kittel_thermal}: 
\begin{equation}
 \mathcal{S}_{ST} =N k_B \, \left[ \frac{5}{2}+\ln \left( \frac{1}{n\lambda_{dB}^3} \right) \right]
\label{ST}
\end{equation}

As early as 1948, Callen~\cite{Callen1948} demonstrated that the Kelvin relation, which can be derived from Onsager reciprocity, implies that the Seebeck coefficient (when it is purely diffusive and not affected by phonon drag) is the ratio of entropy per mobile charge~\cite{Callen1948,Behnia2015b}. It is not surprising, therefore, to see that Eq. \ref{Pisarenko} (with $r=0.5$, which implies a constant scattering time, and an additional factor of 2 due to spin degeneracy) represents the entropy of non-degenerate electrons according to Eq. \ref{ST} per charge carrier.

Above degeneracy temperature, $n \lambda_{dB}^3 <1$ and $\ln \frac{1}{n \lambda_{dB}^3}$ is positive. It quantifies the temperature-dependent entropy of a perfect gas of indiscernible particles when the primitive cell of the phase space is the Planck constant~\cite{Kittel_thermal}. With increasing temperature, $\lambda_{dB}$ shrinks and the number of configurations (for a fixed density of particles) enhances. The heavier the particles, the larger the entropy of the classical gas at a given temperature. This means that the room-temperature entropy of Ar is larger than the room-temperature entropy of Ne~\cite{Panos2015}. In our context of investigation, the same line of reasoning would imply that the heavier non-degenerate electrons, the larger their Seebeck coefficient.

The relevance of Eq.~\ref{Pisarenko} to our data can be seen by plotting the magnitude of the measured Seebeck coefficient at a given temperature as a function of $\ln(n)$. As seen in Fig. \ref{Seebeck}(c), at four different temperatures, our data (in blue) combined with what was reported by previous authors for oxygen-reduced~\cite{frederikse1964} (in green) and La-doped~\cite{Okuda2001,cain2013} (in red) strontium titanate correspond to what is expected according to Eq. \ref{Pisarenko}, with $r=0.5$ and $\Lambda=\lambda_{dB}$. The extracted effective mass is $7 \, m_e$ at 100 K and rises to $13 \, m_e$ at 300 K.

It is instructive to compare the magnitude of the Seebeck coefficient in our system with a common semiconductor such as germanium. As seen in Fig. \ref{Seebeck}(d), at room temperature, the Seebeck coefficient in both Ge and Nb-doped strontium titanate is a linear function of $\ln (n) $. The two lines have identical slopes but are shifted, implying heavier ($13 \, m_e$ in strontium titanate) and lighter ($0.5 \, m_e$ in Ge) carriers. 

\subsection{Temperature dependence of $\Lambda$ and $m^{\star}$}
\begin{figure}[h!]
\centering
\includegraphics[width=\linewidth]{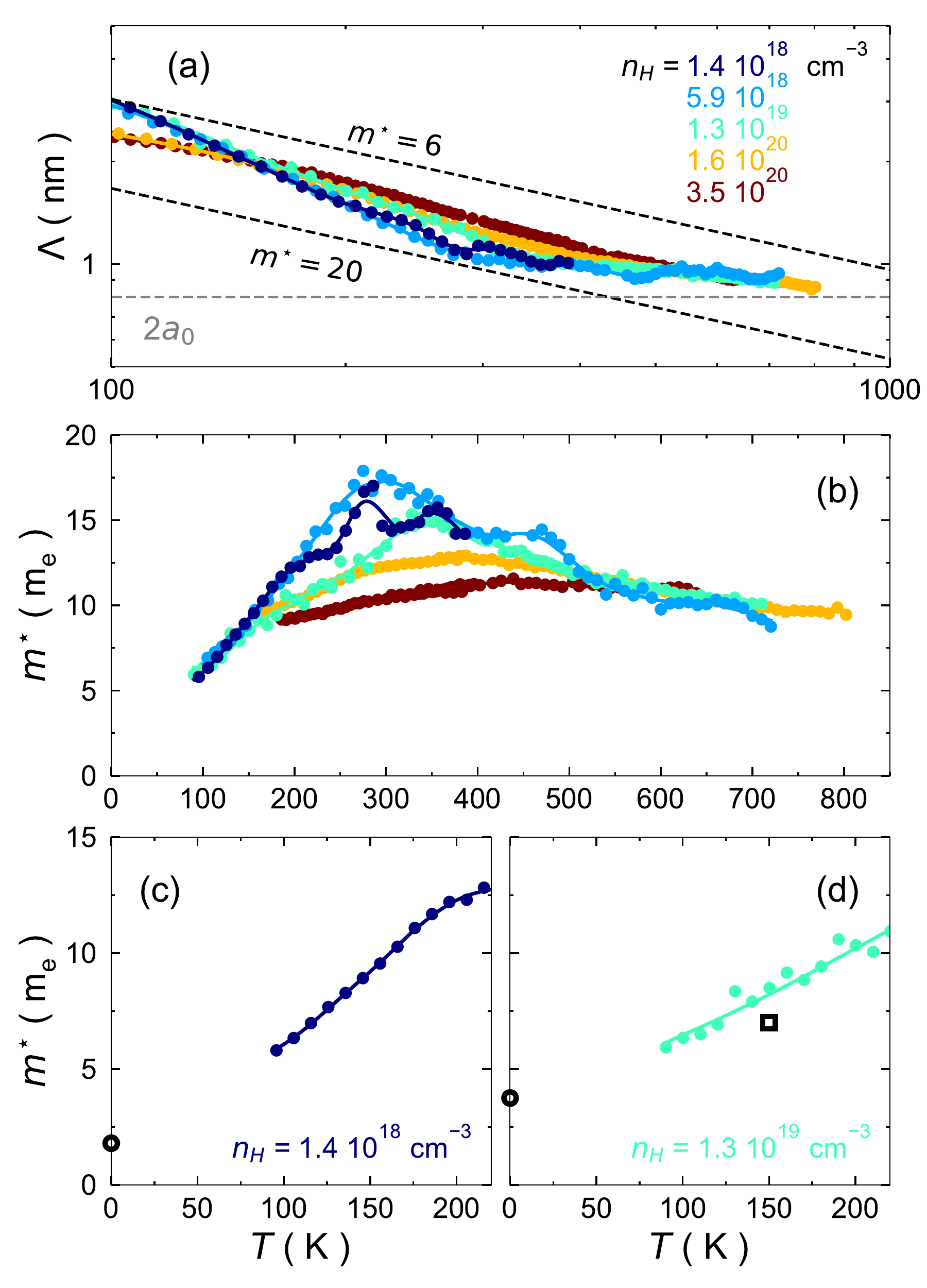}
\caption{\textbf{a)}: Temperature dependency of $\Lambda$ deduced from Eq. \ref{Pisarenko} using the measured Seebeck coefficient [Fig \ref{Seebeck}(a)] and assuming $r=0.5$.
Dashed lines represent the temperature dependence for $\lambda_{dB}(m^{\star}=6~m_e)$ and $\lambda_{dB}(m^{\star}=20~m_e)$. One can see that the data for all temperatures and carrier densities fall between these two lines. \textbf{b)}: Temperature dependence of the effective mass by assuming $\Lambda = \lambda_{dB}$. Note the non-monotonous temperature dependence of $m^{\star}$ and the convergence to a value close to $10~m_e$ at high temperature. \textbf{c)}: The effective mass obtained here compared to the low-temperature effective mass obtained by quantum oscillation below 2 K~\cite{Lin2014} (open circle) at $n_H \simeq 1.4\times 10^{18}$ cm$^{-3}$. \textbf{d)}: The effective mass obtained here compared to the low-temperature effective mass obtained by quantum oscillation below 2 K~\cite{Lin2014} (open circle) and by ARPES at 150~\cite{Chang2010} (open square) at $n_H \simeq 1.3 \times 10^{19}$ cm$^{-3}$.}
\label{Lambda_mstar}
\end{figure}

Fig. \ref{Lambda_mstar}(a) shows the temperature dependence of $\Lambda$, extracted from the Seebeck coefficient combined with Eq. \ref{Pisarenko} and assuming $r=0.5$. If $\Lambda$ is indeed the thermal de Broglie wavelength, its temperature dependence, faster than $T^{-0.5}$ below room temperature, would imply an increasing effective mass. The magnitude of $\Lambda$ would imply a mass between $6 \, m_e$ and $20 \, m_e$ in the entire temperature range and for all five samples.

The effective mass assuming that $\Lambda=\lambda_{dB}(m^{\star})$ is plotted in Fig. \ref{Lambda_mstar}(c). One can see that the extracted mass varies with temperature and with carrier density. The figure also shows that our estimation of high-temperature effective mass displays a reasonable extrapolation to the low-temperature effective masses obtained from quantum oscillations~\cite{Lin2014}. Specifically, when $n_H \simeq 1.5 \times 10^{18}$ cm$^{-3}$, $m^{\star}\simeq 2 ~ m_e $ and when $n_H > 4 \times 10^{18}$ cm$^{-3}$, it passes to $m^{\star}\simeq 4 ~ m_e$~\cite{Lin2014}, as a result of non-parabolic band dispersion of the lower band~\cite{vdMarel2011}. This is in agreement with our extracted masses at $n_H = 1.4 \times 10^{18}$ cm$^{-3}$ and at $n_H =5.8 \times 10^{18}$ cm$^{-3}$. As seen in panel d, our data is also consistent with the $m^{\star} = 7 ~ m_e$ obtained at 150 K from ARPES measurement~\cite{Chang2010}. Note also the non-monotonous evolution of the temperature dependence of $m^{\star}$ as well as the fact that above 500 K, within experimental margin, $m^{\star}\simeq 10 ~ m_e$. 

The reported specific heat data at different doping concentrations and different temperature ranges find a low-temperature mass between $1.8m_e$ and $4m_e$~\cite{Ahrenes2007}. An excellent agreement between specific heat data at optimal doping~\cite{Lin2014_multiple} and quantum oscillation data~\cite{Lin2014} can be obtained~\cite{Collignon2017}, if one assumes that at optimal doping, the lower band is heavier ($m_1= 3.85 \pm 0.35 m_e$)
compared to the higher bands ($m_{2,3}= 1.52 \pm 0.25 m_e$). This is in agreement with what is expected by DFT calculations~\cite{vdMarel2011} and with the magnitude of the measured superconducting penetration depth~\cite{Collignon2017}. As seen above, these values are also in agreement with our extrapolation to lower temperature. 

Note that since the thermal derivative of the entropy of a classical gas does not depend on the mass of the particles, specific heat cannot be used to extract the mass of electrons above the degeneracy temperature. As one can see in Figure 3 of ref. \cite{Panos2015}, in  heavier classical gas, entropy is larger, but its thermal slope is the same.

\subsection{Temperature dependence  of the Bohr radius}
\begin{figure}[h!]
\centering
\includegraphics[width=\linewidth]{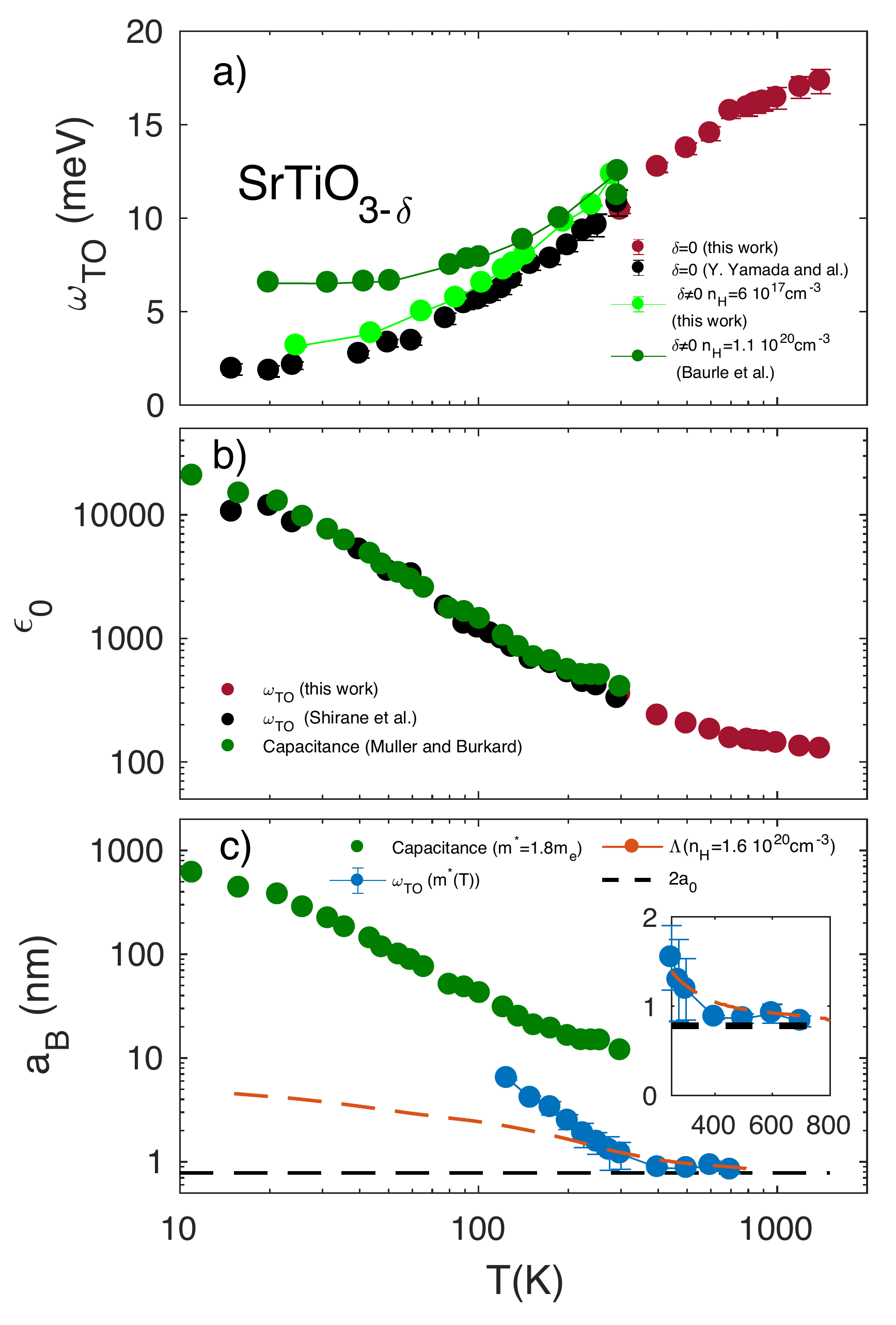}
\caption{\textbf{a):} Temperature dependence of the energy of the soft transverse optical phonon at the zone center extended to 1500 K in insulating SrTiO$_3$ (black and dark red circles) compared to doped SrTiO$_{3-\delta}$ with a carrier density $n_H=6\times 10^{ 17}$cm$^{-3}$ (light green circles) and n$_H=1.1\times 10^{20}$cm$^{-3}$ (dark green circles \cite{Bauerle1980}). \textbf{b):} Temperature dependence of the static electric permittivity $\epsilon$ extracted from the energy of the soft mode (black and dark red circles) compared to what was directly measured~\cite{Muller1979} (dark green circles). \textbf{c):} The effective Bohr radius, combining the temperature-dependent mass extracted from the Seebeck coefficient and the permittivity quantified by neutron scattering data (dark blue) and the de Broglie thermal wavelength extracted from the Seebeck coefficient, $\Lambda$. Also shown is $a_B$ with the low-temperature mass (green circles) neglecting its thermal evolution. As highlighted in the inset, $a_B$ saturates above 400 K to twice the lattice parameter (represented by a dashed horizontal line).}
\label{aB}
\end{figure}

Fig. \ref{aB}(a) shows the temperature dependence of the frequency of the transverse optical phonon, at the center of the Brillouin zone, up to 1500 K obtained from our neutron scattering measurements. Previously, these soft phonons were extensively studied below room temperature by neutons cattering~\cite{Yamada1969b,Bauerle1980}, hyper-Raman spectroscopy~\cite{Vogt95} and optical spectroscopy~\cite{Sirenko2000,vanMechelen2010,Rossle2013}. Our new data, in good agreement with early measurements, shows that the frequency of this mode, which increases from 1-2 meV at 2 K to 11 meV at 300 K~\cite{Yamada1969b,Vogt95}, continues to rise upon warming above room temperature. The figure also shows how the frequency of the soft mode is affected by doping. Above 200 K $\omega_{TO}(T)$ becomes independent of the doping at least up to $n_H = 1.1 \times 10^{20}$ cm$^{-3}$.

In an ionic solid, the static electric permittivity, $\epsilon_0$, and its high-frequency counterpart, $\epsilon_{\infty}$, are linked to the longitudinal and transverse frequencies through the Lyddane-Sachs-Teller relation. When there are multiple longitudinal and transverse optical modes, as in our case, one has~\cite{Cochran1962} :
\begin{equation}\label{5}
\prod_{i=1}^3\frac{\omega_{LO,i}^2}{\omega_{TO,i}^2} =\frac{\epsilon_{0}}{\epsilon_{\infty}}
\end{equation}

Assuming that $\epsilon_{\infty}$ and all other optical modes, other than the soft TO1 phonons do not vary with temperature, this expression implies that $\omega_{TO,1} \propto \sqrt{1/\epsilon_{0}}$ and the magnitude and the temperature dependence of one can be used to track the evolution of the other. Yamada and Shirane~\cite{Yamada1969b} demonstrated that this is indeed the case and $\omega_{TO} \simeq 194.4/ \sqrt{\epsilon_{0}}$, as one can see in Fig. \ref{aB}(b). As discussed in the supplement \cite{SM}, this prefactor is in excellent agreement with the measured values of longitudinal and transverse phonons. We can therefore safely use our data to track the evolution of $\epsilon_{0}$ above room temperature. 

Our data implies that even at a temperature as high as 1500 K, the electric permittivity is two orders of magnitude larger than the vacuum electric permittivity. Nevertheless, compared to its magnitude at 2 K, $\epsilon_{0}$ has dropped by a factor of 200. Combined with the enhancement in $m^{\star}(T)$, this leads to drastic shrinking in the Bohr radius, as illustrated in Fig. \ref{aB}(c). Above 400 K, $a_B$ stops its decrease and saturates to a value of 0.8 nm, almost twice the lattice parameter. Interestingly, this is also the magnitude of $\Lambda$ in this temperature range [see Fig. \ref{Lambda_mstar}(a)]. 

\subsection{Back to the Mott-Ioffe-Regel limit and Planckian dissipation}

\begin{figure}[h!]
\centering
\includegraphics[width=\linewidth]{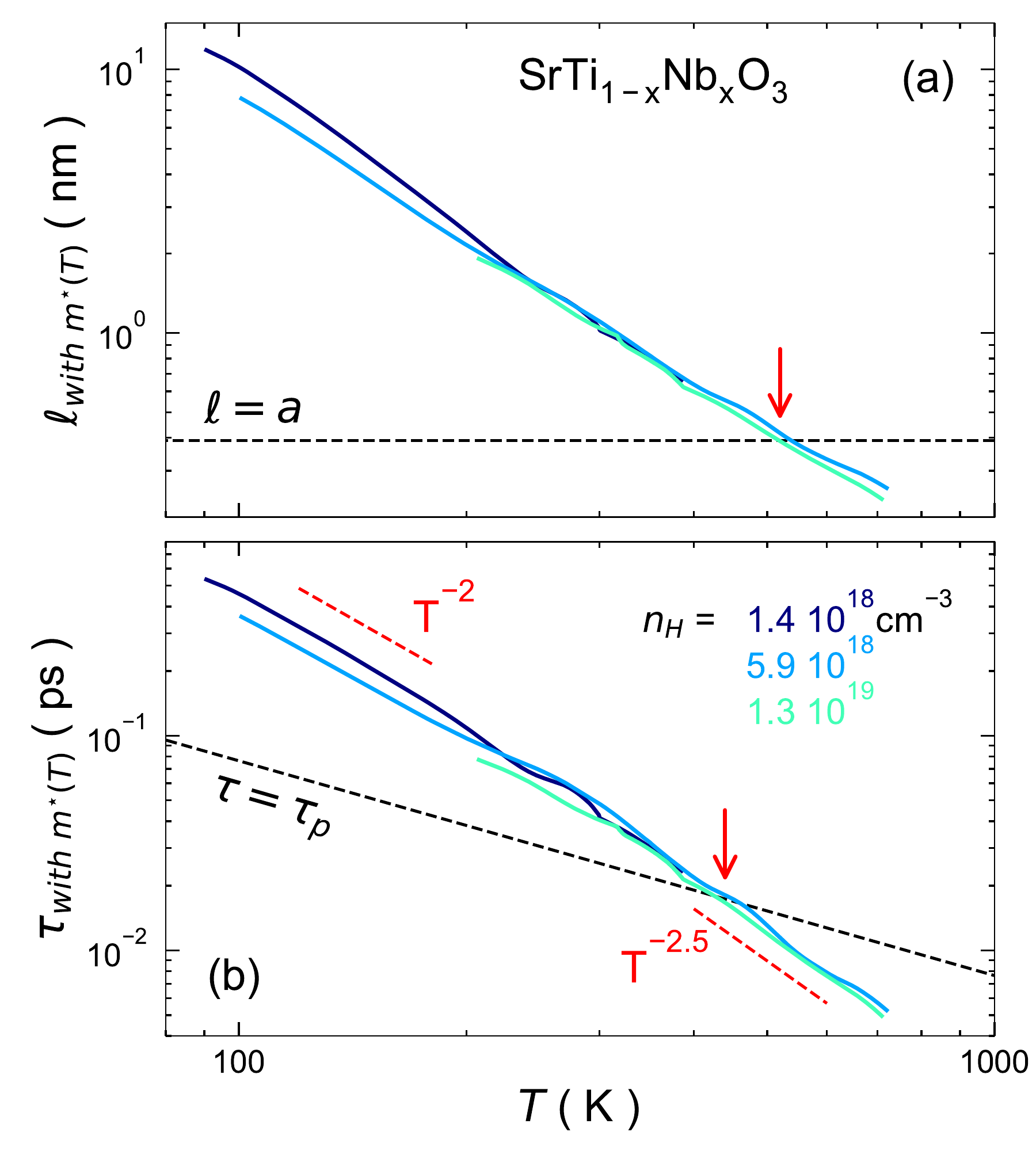}
\caption{
Temperature dependence of the mean free path \textbf{a)} and of the scattering time \textbf{b)} for the three most dilute samples deduced from the resistivity data and the temperature-dependent effective mass obtained from the Seebeck data. With such a temperature dependent mass, the mean-free-path remains longer than the lattice parameter for $T< 520$ K and the scattering time is longer than the planckian time for $T< 450$ K. Both limits are violated at higher temperatures.}
\label{Corrected_MFP}
\end{figure}

Having extracted the temperature dependence of $m^{\star}$, we can return to our resistivity data and compute the mean-free-path and the scattering time in this temperature range taking in to account the temperature dependence of $m^{\star}(T)$. As one can see in Fig. \ref{Corrected_MFP}, both Plankian and MIR limits are verified up to temperatures exceeding room temperature but not above 500 K.

Thus, below 500 K, metallicity of strontium titanate is partially driven by thermal mass amplification. Nevertheless, the mean-free-path remains longer than the interactomic distance and the scattering time exceeds the Planckian time.  This is no more the case above 500 K, i.e. when the thermal wavelength saturates to twice the lattice parameter.

In the non-degenerate regime, the electron velocity is set by the thermal energy: $v_{th} = \sqrt{2k_BT/m^{\star}}$. Therefore, the inequality  $\ell < \Lambda$ is strictly equivalent to the inequality $\tau < \tau_p$ (and vice versa) and a scattering time below the Planckian time means a mean-free-path shorter than the electronic wavelength. The validity of the scattering-based picture in this context becomes questionable.   

\section{Discussion}

\begin{figure}[h!]
\centering
\includegraphics[width=8cm]{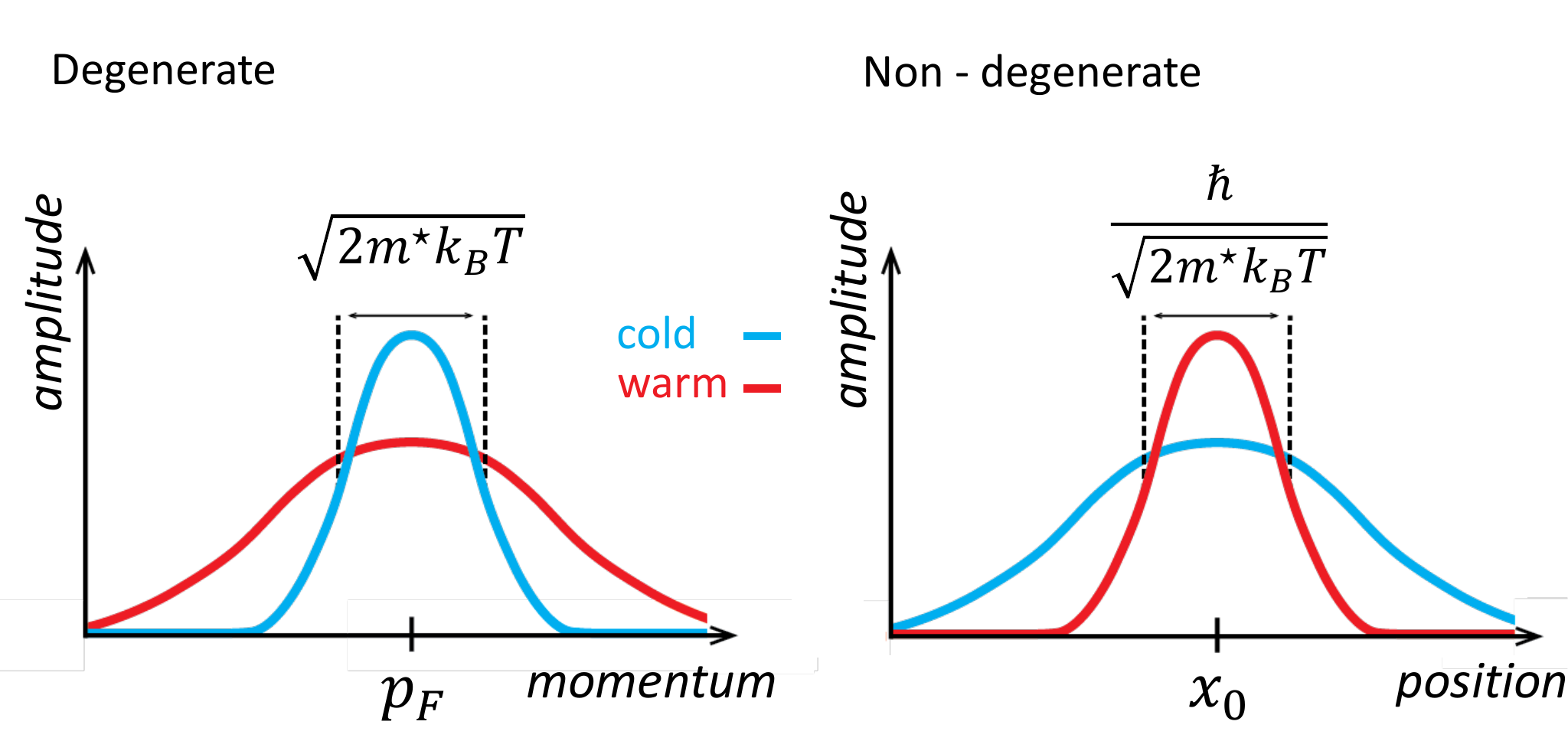}
\caption{\textbf{Left:} In the case of degenerate electrons, heaviness refers to the thermal fuzziness of the Fermi momentum. The heavier the electron the faster the thermal evolution of the sharpness of the amplitude of its wave-functions in the reciprocal space. \textbf{Right:} In the case of non-degenerate electrons, heaviness refers to the thermal fuzziness of the position. The heavier the electron the faster the thermal evolution of the sharpness of the amplitude of its wave-functions in the real space.}
\label{mass}
\end{figure}

\subsection{Mass amplification and its implications}
We saw that our Seebeck data points to a temperature-dependent effective mass. How solid is this conclusion? Can one explain the excess of entropy by a strong energy dependence of the scattering? The answer appears to be negative. Such a route would require an implausibly large $r$ in Eq. \ref{Pisarenko} as discussed in the supplementary (see Section IV of \cite{SM}). There is no independent way to quantify $r$ and ensure that it does not evolve at all with temperature and doping. However, two independent experimental observations establish that $r$ is well below unity and does not evolve with doping. The first is the fact that the Seebeck coefficient at a given temperature is linear in $\ln(n)$, which would have not been the case if $r$ depended on carrier concentration. The second is the fact that above 100 K, mobility at a given temperature shows little or no change with carrier concentration over a very wide window. This implies that the scattering time divided by mass (or alternatively the mean-free-path times the wavelength) is not affected by a shift in the chemical potential. The effective mass has been extracted assuming that $r$ is constant and equal to 0.5, which corresponds to assuming that the scattering time does not depend on carrier concentration. This may not be rigorously true. However, a small change in $r$ would not affect our conclusion that the effective mass is of the order of $10m_e$ at room temperature and above (see \cite{SM} section IV for further details).


This unavoidable heaviness of electrons begs a commentary. In intermetallic solids with f- electrons, heavy quasi-particles are formed upon cooling. This has been extensively documented during the past three decades. In this case, the quasi-particle mass is boosted by accumulation of entropy due to Kondo coupling between localized spin and the Fermi sea~\cite{Hewson}. These are degenerate electrons and the heaviness shows itself in the reciprocal space, where electrons have a Fermi momentum set by the density of carriers. This momentum becomes rapidly fuzzy with warming, leading to large cyclotron masses extracted from temperature dependence of quantum oscillations~\cite{Taillefer1988}. The electron heaviness encountered here implies a process occurring in real space. Non-degenerate electrons have a well-defined position and their momentum is thermal. Warming sharpens this position. A large mass means that the thermal sharpening of this position in real space is unusually fast. Note that in both cases, the mass is an entropy-driven process distinct from the band mass associated with momentum-energy dispersion at a fixed temperature (see Fig. \ref{mass}).

Thus, our conclusion implies a hitherto unknown type of metallicity in a doped polar semiconductor where non-degenerate electrons display a metallic resistivity partially driven by the change in their mass and not merely because of the change in the scattering time. Let us now recall that a temperature-dependent effective mass driven by polaronic effects has been previously suggested in this system.

\subsection{Polarons}

At low temperature, there is a large difference between the static and high-frequency of electric permittivity. This makes our system a natural platform for emergence of polarons. The nature of polarons in SrTiO$_3$ has been subject of numerous theoretical discussions. According to Devreese and co-workers~\cite{Devreese2010}, the optical conductivity data in SrTi$_{1-x}$Nb$_x$O$_3$ can be explained in terms of a gas of large polarons from low to room temperature with no adjustable parameters using a Fr\"ohlich-type interaction with a Fr\"olich-coupling constant $\alpha \simeq2$. This appeared to provide a satisfactory explanation to the less than threefold mass enhancement seen at low temperatures. Indeed, while the expected band mass in 0.7$m_e$~\cite{vdMarel2011}, the experimentally observed mass was 1.8$m_e$~\cite{vanMechelen2010,Lin2013}. Note that in this 'large polaron' picture,  no mass enhancement is expected with rising temperature, because $\alpha$ decreases slightly with rising temperature. 


Many years ago, Eagles argued that the electron mass in doped strontium titanate increases with rising temperature based on the magnitude of the plasma frequency quantified by infrared conductivity measurements~\cite{Gervais1993}. He interpreted the data using a theory of mixed polarons~\cite{Eagles1966,Eagles1969}, where the electronic ground state consists of nearly small polarons and weak-coupling large polarons. In this theory, first applied to Zr-doped SrTiO$_3$ at low temperature~\cite{Eagles1966} and later to SrTi$_{1-x}$Nb$_x$O$_3$ at high temperature~ \cite{Eagles1966,Eagles1969}, the mass enhancement is caused by a change in the electronic overlap integrals or by an increase of the electron-phonon interaction which would increase the contribution of small polarons to the ground state. In the supplementary (see Section II of \cite{SM}), we present a short summary of available reports on the evolution of the plasma frequency by different groups. 
 
 However, the physical basis for such a change in microscopic parameters remains unclear. Indeed, Ciuchi \textit{et al.}~\cite{Ciuchi2000} have demonstrated that the radius and the mass of Fr\"olich-type polarons both decrease with increasing temperature for a large range of coupling constant values. In the words of Fredrikse and co-workers~\cite{frederikse1964}, the polaron ends up 'undressing' with warming in any polaronic picture. Therefore, a quantitative account of temperature-induced mass amplification in a polaron-based picture is missing. Taken on its face value, this implies that the polarons are 'dressing up' with warming (instead of 'dressing down').

\begin{figure}[h!]
\centering
\includegraphics[width=8cm]{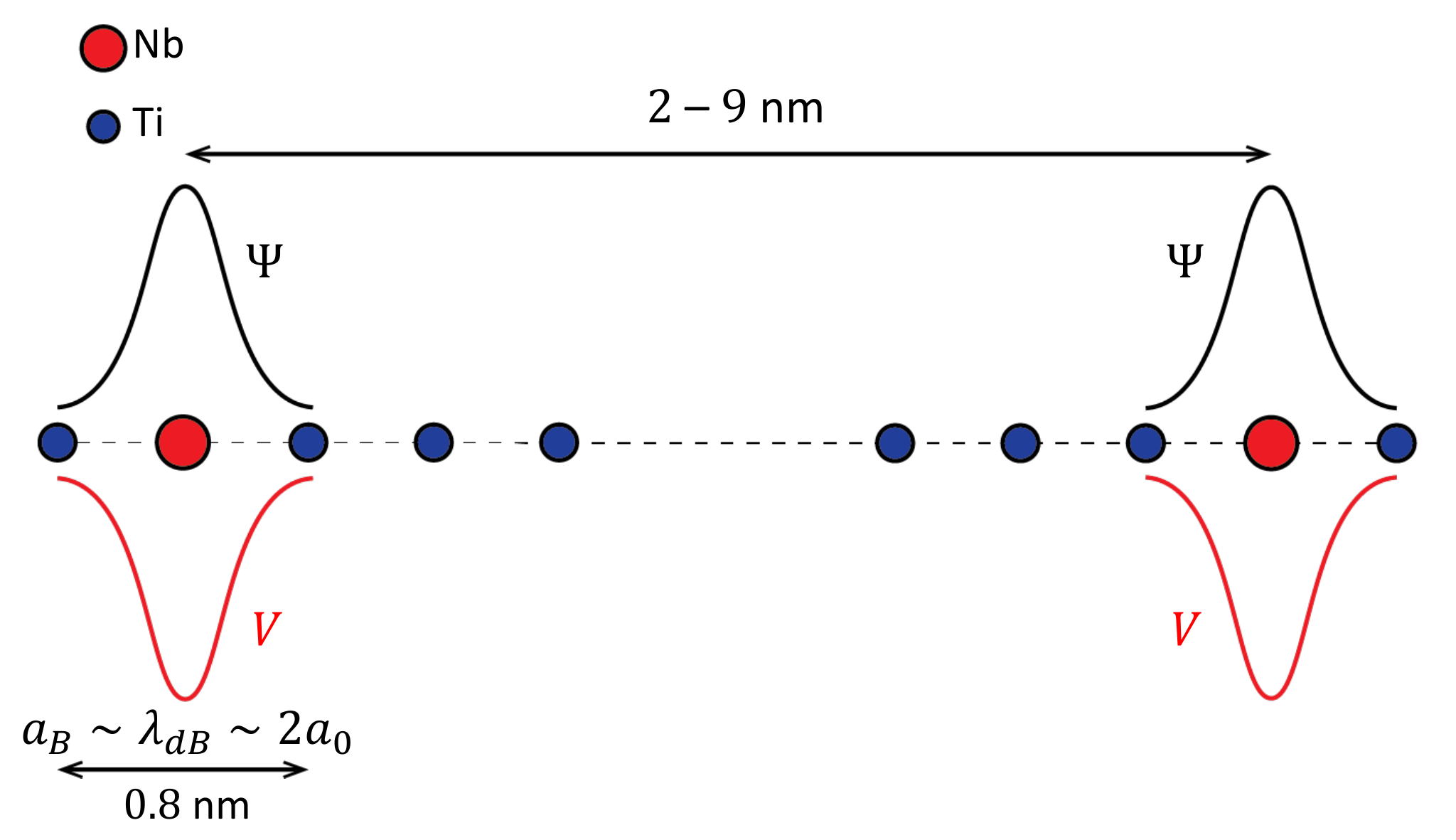}
\caption{A sketch of electron wave-function and the electrostatic potential digged by substituting Ti with Nb. Above 500 K, the Bohr radius and the de Broglie wavelength both shrink to twice the lattice parameter (0.8 nm). The interdopant distance is significantly longer. Yet, a finite metallic conductivity survives.}
\label{L_scales}
\end{figure}

\subsection{Metallicity above 500 K}
We saw that above room temperature, the effective mass ceases to increase, because $\Lambda$, from which it is extracted, saturates above 500 K. In this regime, even with the amplified mass, the  mean-free-path of charge carriers is below the distance between neighboring atoms and shorter than the wavelength of electrons. This is a challenge for any transport picture based on quasi-particle scattering.   In this temperature range, the Bohr radius also saturates, such that $a_B \simeq \Lambda \simeq 2 ~ a_0$ (see Fig. \ref{L_scales}). Nevertheless, the extracted mean-free-path continues to shrink and as a consequence metallicity persists. 

The finite conductivity and its decrease with warming can be formulated in Landauer's picture of conduction viewed as transmission~\cite{Imry1999}. One can state that along 'wires' connecting adjacent dopant sites transmission remains finite, but smoothly decreases with warming. In such a picture, there is no need to invoke quasi-particles and their scattering before their proper formation.

The expression for the Seebeck coefficient used to extract the effective mass assumes a classical gas where particle permutation is allowed. Therefore, electrons are not immobilized at the dopant sites. They are strongly coupled to phonons. As seen in  Fig. \ref{Corrected_MFP}, the scattering time follows T$^{-2.5}$ near 500 K. Let us recall that in a conventional picture of electron-phonon scattering, when non-degenerate electrons are scattered by phonons above their Debye temperature the expected behavior for scattering time is T$^{-1.5}$. To sum up, the sheer magnitude of the mean-free-path is problematic in a scattering based picture and its temperature dependence is faster than what would have been expected in a  familiar context. 

It is instructive to compare the thermal energy of electrons and the depth of the Coulomb potential well. At 500 K, $\epsilon_{0}= 150$ and $\epsilon_{\infty}=6$. The dynamic and the static Coulomb energy at a Bohr radius are respectively $V_{0}= 4.5$ meV  and $V_{\infty}= 102$ meV. The kinetic energy at 500 K (43 meV) exceeds $V_{0}$, which is low thanks to the screening provided by soft phonons. However,  the time scale for thermal electrons($\hbar/k_BT$) is shorter than the time scale for soft phonons ($\hbar/\omega_0(T)$) when T$\sim$ 500 K and therefore this screening may be too slow for faster electrons. A proper treatment of this issue  requires proper documentation of the frequency and the wave-vector dependence of the electric permittivity. 

\section{Summary}
This paper presents extended measurements of resistivity and the Seebeck coefficient of Nb-doped strontium titanate to very high temperatures. We started by showing that if one assumes a temperature-independent effective mass, then the very low magnitude of the mobility implies a mean-free-path and a scattering time too short compared with the lowest plausible values. We then showed that the magnitude, the doping dependence and the thermal evolution of the Seebeck coefficient imply a revision of the mean-free-path and scattering time. The extracted effective mass extrapolates smoothly to the mass obtained at 2 K by quantum oscillations and at 150 K by ARPES~\cite{Chang2010}. Injecting this temperature-dependant effective mass to the analysis of resistivity allows us to correct our estimation of the mean free path and of the scattering time finding that both the MIR and the Planckian limit are respected at room temperature, but not above 500 K. 

In the Drude picture, the resistivity of a metal increases with warming, because a fixed number of carriers scatter more frequently with rising temperature. In the case of doped strontium titanate, this picture needs a serious correction, because the enhanced resistivity is partially driven by mass amplification. To the best of our knowledge, there is no microscopic theory for this behavior. However, the similarity between doped strontium titanate and other dilute metals near an aborted ferroelectric order raises the suspicion that such a behavior is intimately connected with presence of a soft ferrolectric mode. For temperatures exceeding 500 K, the magnitude of the extracted mass is not sufficiently large to impede the violation of the expected boundaries. Metallicity persists even when the distance between two scattering events is shorter than what is needed to make the existence of a  charge carrier meaningful. This is  a stark case of metallicity 'beyond quasi-particles'~\cite{Lin2017,mishchenko2019,Zhou2019}.

\section*{Methods}
We measured the electric resistivity of six and the Seebeck coefficient of five niobium doped strontium titanate samples commercially obtained from Crystec with Nb content varying between 0.02 and 2 atomic percent. The samples had approximate dimensions of $2.5 \times 5 \times 0.5$ mm$^{3}$. For all measurement electrical contacts are made with thermal evaporation of gold and 4029 Dupont silver paste. Inelastic neutron scattering experiment have been performed at Orph\'ee reactor. Details are given in the supplementary (see Section V of \cite{SM}).

The resistivity, Hall and Seebeck measurements from 2 to 400 K was done in a Quantum Design Physical Property Measurement System (PPMS). The carrier concentration was found to be almost constant between 2 and 300 K as previously reported. The Seebeck coefficient was measured with an usual "heat on / heat off" technique. One extremity of the sample is glued with silver paint to a copper block that act as a thermal drain, the temperature gradient is applied via a RuO$_2$ resistor and the thermal gradient is measured thanks to two type-E thermocouples. The thermal gradient was kept below 10\% of the average temperature of the sample.

Electrical resistivity (up to 900 K) and the Seebeck coefficient (up to 800 K) were both measured with a custom-made probe using a 50W Watlow heater. Sample temperature and thermal gradient were monitored by  Pt-100 thermometers directly glued on the sample with 6038 Dupont silver paste. To avoid any additional doping through oxygen reduction,  measurements were performed under a rough vacuum of about 10$^{-2}$ mbar, well below what would  dope pristine samples. No hysteresis was found upon heating and cooling the sample The data found in the low-temperature and high-temperature sweeps were found to overlap with a reasonable multiplicative factor of the order of 10\% due to a difference in the geometrical factor between the two setups.

\section*{Acknowledgments}

We thank S. Fratini, A. Georges, S. A. Hartnoll, J. Hemberger, H. Kang, S. A. Kivelson, X. Lin, T. Lorenz, D. Maslov, A. Millis, J. Mravlje, C. W. Rischau and J. Ruhman for stimulating discussions. We thank M. Delbecq for his help in the early days of this project. This work is supported by the Agence Nationale de la Recherche (ANR-18-CE92-0020-01) and by Jeunes Equipes de l$'$Institut de Physique du Coll\`ege de France.

\sloppy
\bibliography{biblio}
\bibliographystyle{apsrev4-1}

\pagebreak
\begin{center}
\textbf{\large Supplemental Material}
\end{center}
\setcounter{equation}{0}
\setcounter{figure}{0}
\setcounter{table}{0}
\setcounter{section}{0}
\setcounter{page}{1}
\makeatletter
\renewcommand{\thesection}{S~\Roman{section}}
\renewcommand{\theequation}{S\arabic{equation}}
\renewcommand{\thefigure}{S\arabic{figure}}

\section{The exponent of the power law in the temperature dependence of resistivity}
\label{exponent_annex}
\begin{figure}[h!]
\centering
\includegraphics[width=\linewidth]{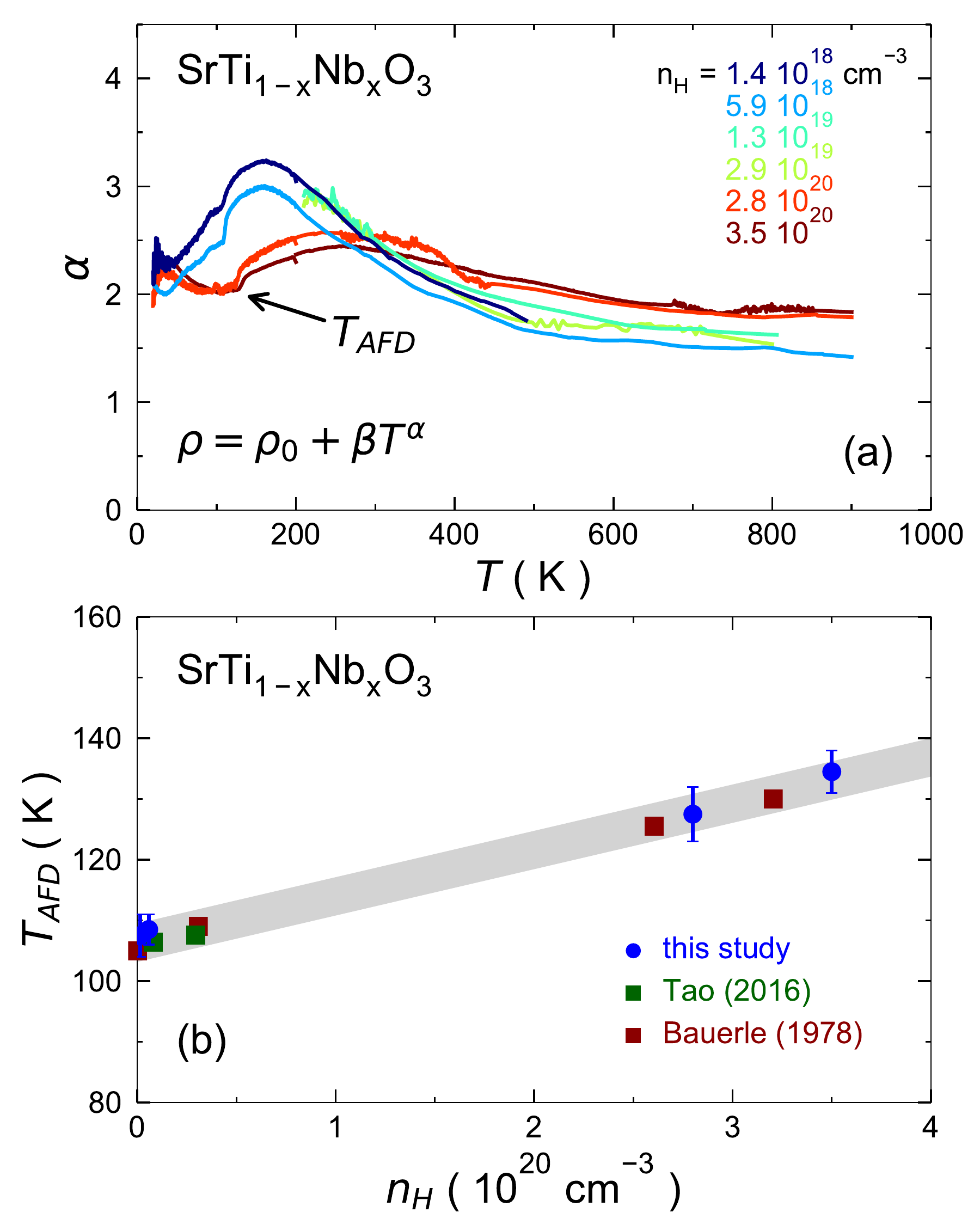}
\caption{
\textbf{a)}: Temperature dependency of the exponent $\alpha$ of the resistivity: $\rho = \rho_0 + \beta T^{\alpha}$. $\alpha$ is extracted from the following derivative: $\alpha = \partial \ln ( \rho-\rho_0 ) / \partial \ln T$.
The kink in $\alpha$, emphasised by the black arrow, is due to the antiferrodistortive transition as previously discussed in \cite{Lin2017}.
\textbf{b)}: Critical temperature of the antiferrodistortive transition as a function of doping, extracted from the anomaly in $\alpha$ reported pannel a). We also reproduce $T_{AFD}$ extracted from sound velocity measurements \cite{Bauerle1978} and Montgomery measurements \cite{Tao2016}. The three sets of data are in quantitative agreement.}
\label{exponent}
\end{figure}

Fig. \ref{exponent}(a) shows the temperature dependence of the exponent $\alpha$, assuming a simple power-law behavior for resistivity: $\rho=\rho_0+ T^\alpha$. We used $\alpha = \partial \ln ( \rho-\rho_0 ) / \partial \ln T$ to extact $\alpha$.

As one can see in the figure, the much-discussed quadratic temperature dependence of resistivity~\cite{Okuda2001,vdMarel2011,Lin2015sc,mikheev2015,mikheev2016,Maslov2016} is a low-temperature phenomenon. The exponent of resistivity is not constant and evolves with temperature, as previously reported~\cite{Lin2017,collignon2019metallicity}. Below 40 K, $\alpha \simeq 2$. Above this temeprature, it shows a non-monotonous evolution with rising temperature and near room-temperature, $\alpha \simeq 3$. According to the new data extended to high temperature , the exponent continues to decrease up to 900 K. The resistivity is still far from linear in temperature with $\alpha \simeq 1.7$.

As indicated by the arrow Fig. \ref{exponent}(a), the temperature dependence of the exponent of the power law of the resistivity shows an anomaly, at a temperature which depends on the doping. As noticed before~\cite{Lin2017}, this anomaly occurs at the temperature range ($105$ K $<T< 140$ K) of the cubic-tetragonal structural transition. In this antiferrodistortive transition, neighboring octahedra tilt in opposite orientations.
As seen in Fig. \ref{exponent}(b), the extracted critical temperature $T_{AFD}$ of our Nb-doped samples is in quantitative agreement with previous studies using sound velocity~\cite{Bauerle1978} or the Montgomery technique probing resistivity anisotropy~\cite{Tao2016}.

\section{Plasma frequency and mass enhancement}
\label{Omegap_Annex}
\begin{figure}[h!]
\centering
\includegraphics[width=\linewidth]{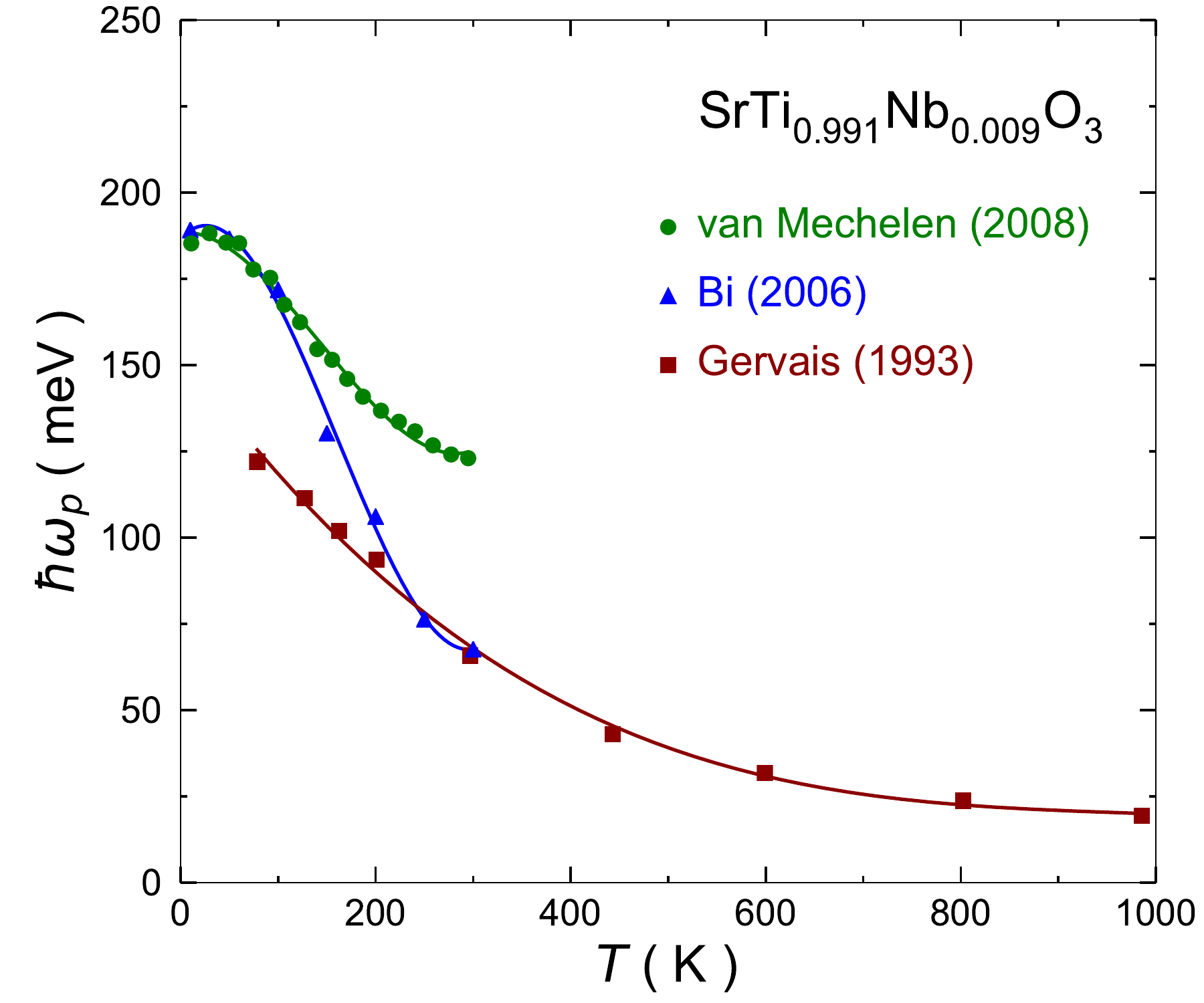}
\caption{Temperature dependency of the plasma frequency $\omega_p$ extracted from a fit of the reflectivity spectrum by Gervais \textit{et al.} \cite{Gervais1993} and Bi \textit{et al.} \cite{bi2006} or by a fit of the Drude peak by van Mechelen \textit{et al.} \cite{vMechelen2008}.
All three measurement are carried at the same Nb doping $x=0.009$ ($n\sim 1.5 ~ 10^{20}$ cm$^{-3}$).
}
\label{plasma_freq}
\end{figure}

As discussed in the main text, the plasma frequency quantified by infrared conductivity was invoked by Eagles \textit{et al.} \cite{Eagles1996} to argue that polarons in strontium titanate become heavier upon warming. The data used for this scenario was reported by Gervais \textit{et al.}~\cite{Gervais1993}. It is instructive to compare these results with what was reported by two other groups.

Bi \textit{et al.}~\cite{bi2006} fitted their reflectivity spectrum to obtain the plasma frequency. On the other hand, van Mechelen \textit{et al.} \cite{vMechelen2008} deduced the same quantity by fitting the Drude peak in the conductivity spectrum. As seen in Fig. \ref{plasma_freq}, the three sets of measurements do not quantitatively agree with each other. However, they all exhibit a clear decrease in $\omega_p$ with increasing temperature. Note that while van Mechelen \textit{et al.} report the bare plasma frequency, the two other groups report the screened one. Therefore, a factor of $\sqrt{\epsilon_{\infty}} = 2.3$ is to be applied to compare that data. 

The plasma frequency, $\omega_p^2$, is proportional to the inverse of the effective mass:
\begin{equation}
\omega_p^2 = \frac{ne^2}{\epsilon_{\infty} m^{\star}}
\end{equation}

Therefore, the decrease in plasma frequency with warming implies a concomitant increase in the effective mass. This is in qualitative agreement with the conclusions of the present paper, at least for temperature below 400 K. Future optical measurements are desirable to pin down the link between our DC transport data and high-frequency conductivity.

\section{Derivation of the 'Pisarenko' formula}
\label{Pisarenko_annex}
In this section, following the reasoning of Ioffe \cite{Ioffe1957}, we show how the 'Pisarenko' formula can be derived in a simple picture. Note however, that the same equation could be derived by assuming that i) electrons belong to a classical gas  with Sackur-Tetrode entropy~\cite{Kittel_thermal}; and ii) the Seebeck coefficient quantifies the entropy flow per traveling charge carrier in absence of thermal gradient~\cite{Callen1948,Behnia2015b}. 

Let us consider a current of electrons $j \propto \int^{\infty}_0 j(\epsilon) d \epsilon$ so that the average energy of carriers electrons is:
\begin{equation}
 <\epsilon> = \frac{\int^{\infty}_0 \epsilon j(\epsilon) d \epsilon}{\int^{\infty}_0 j(\epsilon) d \epsilon}
\label{s1}
\end{equation}
With $f_0$ the energy distribution function of our electrons and $\ell$ their mean free path we can rewrite the current as:
\begin{equation}
 j \propto \int^{\infty}_0 \frac{\partial f_0}{\partial \epsilon} \epsilon \ell ( \epsilon ) d \epsilon
\label{s2}
\end{equation}
We assume a mean free path depending on the energy as $\ell \propto \epsilon ^{r}$ (which is equivalent to the $\tau \propto \epsilon ^{r+1/2}$ assumption made in the text).
Now we consider the degenerate regime, therefore the equilibrium energy distribution function is the Maxwell-Boltzmann distribution $f(\epsilon)=e^{-\frac{\epsilon-\mu}{k_B T}}$, where here $\mu$ is the chemical potential.
By using equations \ref{s1} and \ref{s2} we deduce:
\begin{equation}
 <\epsilon> = \frac{\int^{\infty}_0 \epsilon^{r+3} e^{-\frac{\epsilon}{k_B T}} d \epsilon}{\int^{\infty}_0 \epsilon^{r+2} e^{-\frac{\epsilon}{k_B T}} d \epsilon}
\label{s3}
\end{equation}
And with an integration by parts of the numerator of equation \ref{s3} we get:
\begin{equation}
 <\epsilon> = k_BT \, (2+r)
\label{s4}
\end{equation}
As the Seebeck coefficient, $S$, is the entropy, $\mathcal{S}$, per carrier we get:
\begin{equation}
 S=\frac{\mathcal{S}}{e}=\frac{1}{e}\frac{<\epsilon>-\mu}{T}=\frac{k_B}{e} \left( 2+r - \frac{\mu}{k_BT} \right)
\label{s5}
\end{equation}
As pointed out by Okuda \textit{et al.} in the same context \cite{Okuda2001}, the chemical potential can be deduced by considering a temperature independent carrier concentration:
\begin{equation}
 n = z \int^{\infty}_0 \mathcal{D}(\epsilon)f(\epsilon)d\epsilon
\label{s6}
\end{equation}
With $z$ the degeneracy (namely 2 because of the electron spin) and $\mathcal{D}(\epsilon)$ the density of states.
Once again assuming a Maxwell-Boltzmann statistic, we get from equation \ref{s6}:
\begin{equation}
 n = 2 \lambda_{dB}^{-3} e^{\frac{\mu}{k_B T}} \Leftrightarrow \mu = k_B T \, \ln \left( \frac{n \lambda_{dB}^3}{2} \right)
\label{s7}
\end{equation}

Equations \ref{s7} and \ref{s5} give us the Pisarenko formula:

\begin{equation}
 |S| = \frac{k_B}{e} \left[ 2+r - \ln \left( \frac{n\lambda_{dB}^3}{2} \right) \right]
\label{Pisarenko2}
\end{equation}

\begin{figure}[h!]
\centering
\includegraphics[width=\linewidth]{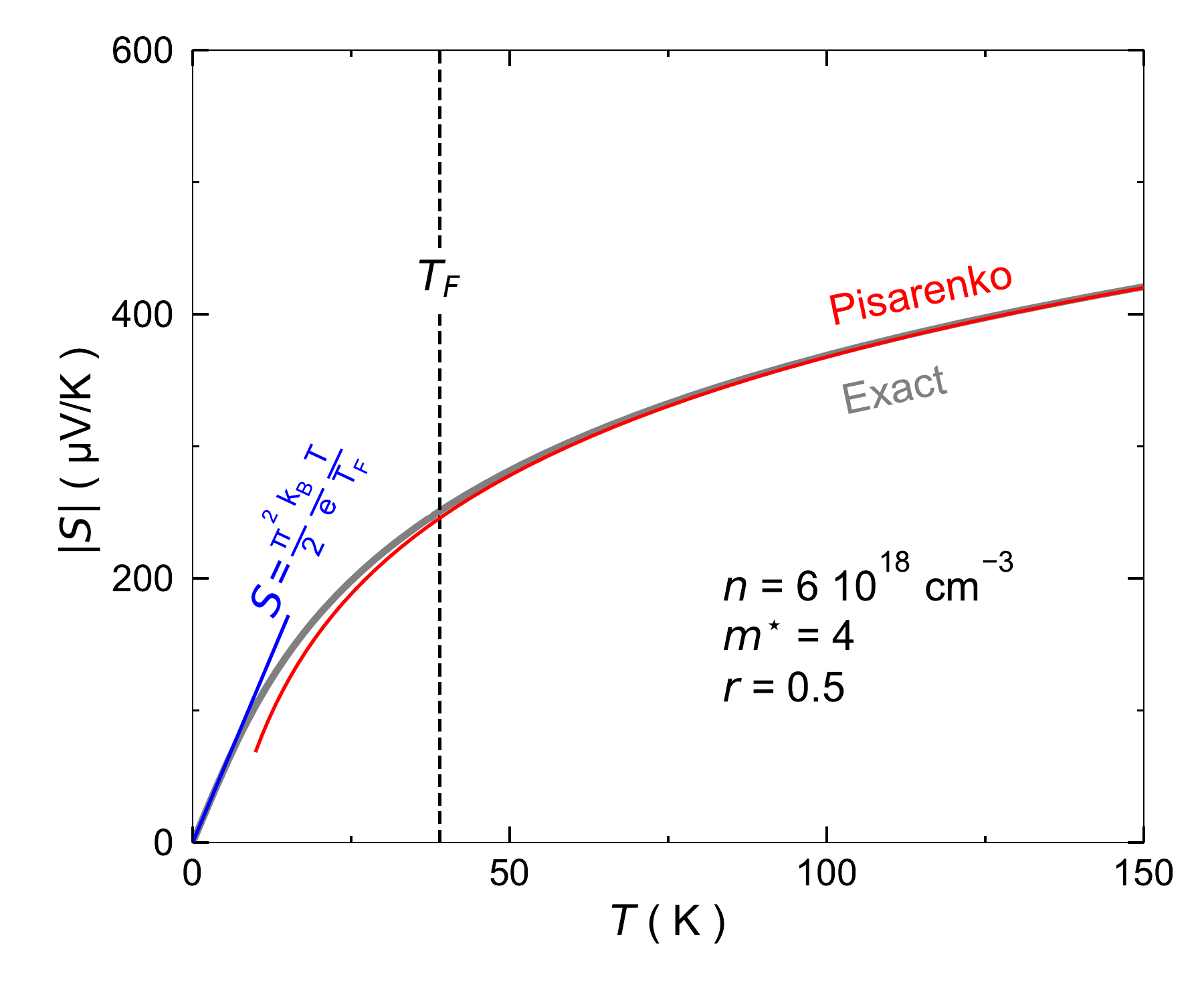}
\caption{Seebeck coefficient seen via different formulae: the low temperature approximation (in blue), the Pisarenko high temperature approximation (in red) computed from equation \ref{Pisarenko2} and the exact derivation (in grey). The Pisarenko formula becomes valid with negligible error at the Fermi temperature $T_F$. Computation were done assuming a metal with $n_H = 6 \times 10^{18}$ cm$^{-3}$, $m^{\star} = 4 ~ m_e$ and a scattering parameter $r=0.5$ as stated on the figure.
}
\label{Pisarenko_supp}
\end{figure}

A comparison between the numerically computed exact formula for the Seebeck coefficient (\textit{i.e.} using the Fermi Dirac distribution for electrons at all temperatures) and the Pisarenko approximation is shown in figure \ref{Pisarenko_supp}.
Above the Fermi temperature, the error on $S$ becomes negligible.

\section{Scattering parameter \textit{r}}
\label{ScatteringParam_annex}

In the main text, we assumed that $r=0.5$ in Eq. 4 for any temperature and doping. Let us see what would happen if one assumes that the effective mass remained constant and the evolution of the Seebeck coefficient was driven by variation of $r$.

\begin{figure}[h!]
\centering
\includegraphics[width=\linewidth]{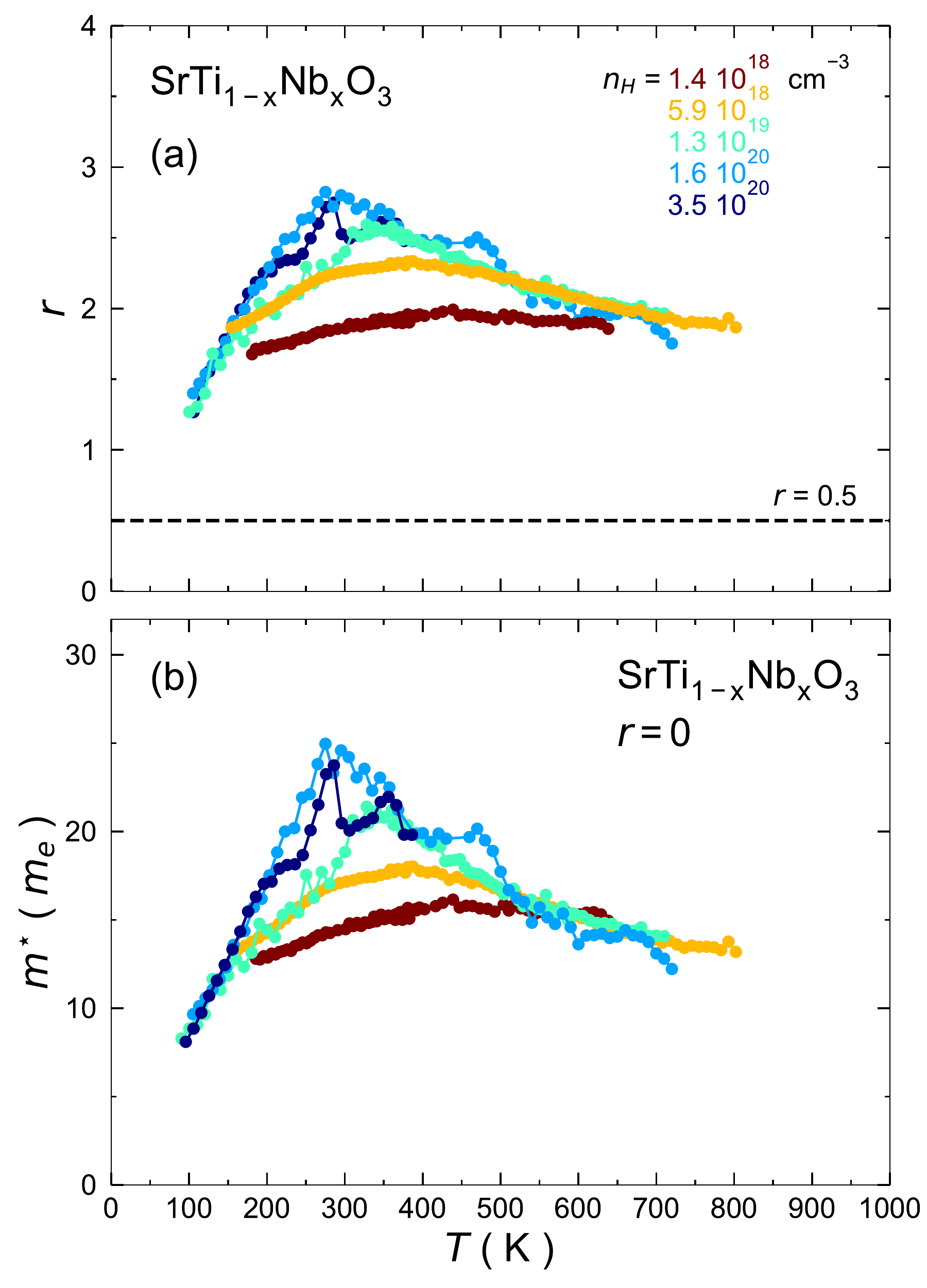}
\caption{
\textbf{a)}: Temperature dependence of the scattering parameter $r$ ($\ell \propto \epsilon^r$) extracted from the Seebeck coefficient reported in Fig.4 (a) in the main text, thanks to the Pisarenko equation (see Eq. \ref{Pisarenko2}) and assuming a temperature independent effective mass, $m^{\star} = 3.8$.
The large value of $r$, several times larger than $r=0.5$ seems improbable, as it supposed a scattering time decreasing with increasing energy.
\textbf{b)}: Temperature dependence of the effective mass extracted from the measured Seebeck coefficient, assuming this time that $r=0$.
}
\label{scattering_parameter}
\end{figure}

Fig. \ref{scattering_parameter}(a) shows $r=\frac{e}{k_B} |S| -2 +\ln{(n \lambda_{dB}^3 / 2)}$ as a function of temperature, assuming that the effective mass is constant and equal to $m^{\star}=3.8$. One can see that that in that case, the Seebeck data would imply a large and variable $r$, much larger than unity. However, this is implausible, because it would imply a mean-free-path with a superlinear energy dependence. Since $\tau \propto E^{r-1/2}$, if $r>0.5$, the scattering time will be longer for electrons with higher energy. 

Let us note that the fact that mobility is constant even when the carrier density changes by many orders of magnitude indicates that scattering time or mean-free-path do not respond significantly to a drastic shift in the chemical potential. In the case of germanium, Johnson and Lark-Horovitz fit the Seebeck coefficient using $r=0$~\cite{johnson1953}, which implies an energy-independent mean-free-path. In Fig. \ref{scattering_parameter}(b), we show how the mass would change in the case of STO assuming $r=0$.



\section{Neutron scattering experiments}
\label{INS}

\begin{figure}[h!]
\centering
\includegraphics[width=0.8\linewidth]{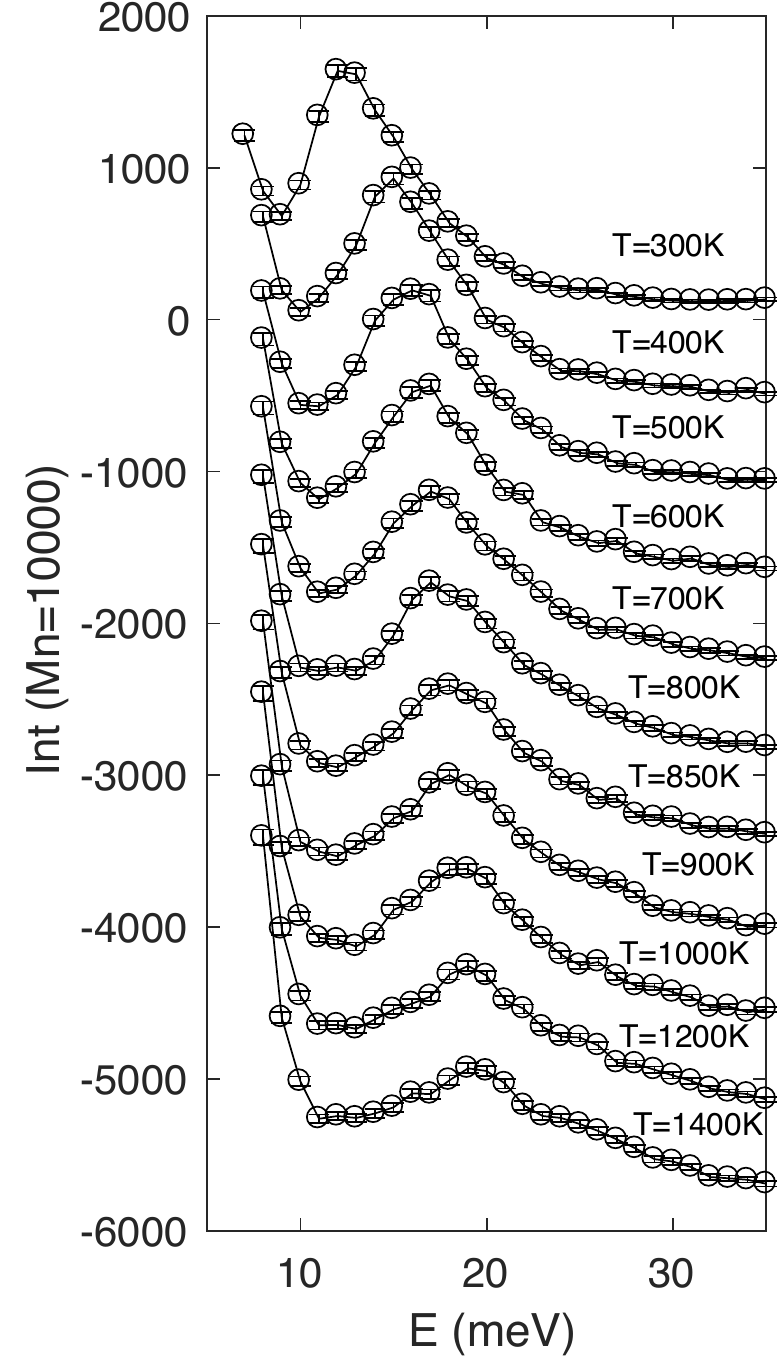}
\caption{Energy scans between 5 to 35 meV at $Q=(2,0,0)$ from $T=300$ K to $1400$ K for S$_2$. Curves are shifted for clarity. As temperature increases, the energy of the TO-mode shifts to higher energy. Energy scans have been fitted a damped harmonic oscillator convoluted with the spectrometer resolution. The temperature dependence of the energy peak position deduced from the fits is reported on Fig. 6 in the main text.}
\label{INS_fig}
\end{figure}

In order to track the evolution of the TO-mode at the zone center we conducted inelastic neutron scattering (INS) experiments in two SrTiO$_3$ samples: S$_1$ ($a=5 \times 5 \times 5$ mm$^3$ reduced sample of $n_H = 6$ $10^{17}$ cm$^{-3}$ from SurfaceNet) and S$_2$ ( $a \approx 25\times25\times15$ mm$^3$ undoped sample from Crystec). Both samples have a mosaicity of $\sim 1 ^{\circ}$. Measurements below 300 K on S$_1$ have been conducted on the 4F2 triples axis spectrometer mounted on a cold beam located at the Orph\'ee reactor in Saclay while measurement on S$_2$ from 300 K up to 1400 K have been conducted on the 2T triple-axis spectrometer installed on the Orph\'ee's thermal beam. For both experiments, the samples were mounted on the scattering plane (100)/(011) with an incident neutron beam of energy $E_i =14.7$ meV. Two PG filters were inserted on the scattered beam on 2T in order to eliminate double scattering. We report on Fig.\ref{INS_fig} the energy scan at the zone center $Q=(2,0,0)$ for sample S$_2$. The temperature dependence of the energy position of the TO-mode for the samples S$_1$ and S$_2$ are reported on Fig. 6 in the main text and compare well with early investigations \cite{Yamada1969b}. 

\section{Lyddane-Sachs-Teller relation and the measured values of optical phonon modes}

As discussed in the main text, the Lyddane-Sachs-Teller (LST) relation in a solid with several atoms links optical and transverse optical modes. Following Yamada and Shirane~\cite{Yamada1969b} we use the expression $\omega_{TO} \simeq 194.4/ \sqrt{\epsilon_{0}}$, to extract $\epsilon_{0}$ in Fig. 6(b) and to quantify the Bohr radius in Fig. 6 (c) of the main text. Let us compare this prefactor with the measured values of longitudinal and transverse phonons. The infrared conductivity measurements by van Mechelen \textit{et al.} led to the following values for the frequency of the three longitudinal optical modes: $\Omega_{LO1}$ = 21.2 meV; $\Omega_{LO2}$ = 58.4 meV; $\Omega_{LO3}$ = 98.7 meV~\cite{vanMechelen2010}. The same measurements found for the two rigid transverse optical modes: $\Omega_{TO2}$ = 21.2 meV; $\Omega_{TO3}$ = 58.4meV~\cite{vanMechelen2010}. We note that the energy position of these three LO modes and two TO modes are independent of the temperature and the carrier density~\cite{vanMechelen2010}. On the other hand the soft TO mode was found to vary from $\Omega_{TO1}$(300 K) = 11.5 meV at room temperature to $\Omega_{TO1}$(7 K) = 2.23 meV at low temperatures, in agreement with neutron scattering \cite{Shirane1969}. Combining these with the $\epsilon_{\infty}\simeq 5$, the LST relation yields:
\begin{equation}
\label{6}
\omega_{TO3} = \frac{\sqrt{5}}{\sqrt{\epsilon_{0}}}\frac{98.7 \times 58.4 \times 21.2}{67.6 \times 21.2}=\frac{191}{\sqrt{\epsilon_{0}}}
\end{equation}
This prefactor, with an accuracy of two percent is equal to the proportionality found by Yamada and Shirane~\cite{Yamada1969b} between the quantities measured by two different techniques and employed in Fig.6.

\end{document}